
%
%
%
\def\unredoffs{} \def\redoffs{\voffset=-.31truein\hoffset=-.48truein}
\def\speclscape{}
%
%
%
%
%
\newbox\leftpage \newdimen\fullhsize \newdimen\hstitle \newdimen\hsbody
\tolerance=1000\hfuzz=2pt
\catcode`\@=11 
\ifx\hyperdef\UNd@FiNeD\def\hyperdef#1#2#3#4{#4}\def\hyperref#1#2#3#4{#4}\fi
\def\bigans{b }
\def\answ{b }
%
\ifx\answ\bigans\message{(This will come out unreduced.}
\magnification=1200\unredoffs\baselineskip=16pt plus 2pt minus 1pt
\hsbody=\hsize \hstitle=\hsize 
\else\message{(This will be reduced.} \let\l@r=L
\magnification=1000\baselineskip=16pt plus 2pt minus 1pt \vsize=7truein
\redoffs \hstitle=8truein\hsbody=4.75truein\fullhsize=10truein\hsize=\hsbody
\output={\ifnum\pageno=0 
  \shipout\vbox{\speclscape{\hsize\fullhsize\makeheadline}
    \hbox to \fullhsize{\hfill\pagebody\hfill}}\advancepageno
  \else
  \almostshipout{\leftline{\vbox{\pagebody\makefootline}}}\advancepageno
  \fi}
\def\almostshipout#1{\if L\l@r \count1=1 \message{[\the\count0.\the\count1]}
      \global\setbox\leftpage=#1 \global\let\l@r=R
 \else \count1=2
  \shipout\vbox{\speclscape{\hsize\fullhsize\makeheadline}
      \hbox to\fullhsize{\box\leftpage\hfil#1}}  \global\let\l@r=L\fi}
\fi
%
\newcount\yearltd\yearltd=\year\advance\yearltd by -2000

\def\Title#1#2{\nopagenumbers\abstractfont\hsize=\hstitle\rightline{#1}%
\vskip 1in\centerline{\titlefont #2}\abstractfont\vskip .5in\pageno=0}
\def\Date#1{\vfill\leftline{#1}\tenpoint\supereject\global\hsize=\hsbody%
\footline={\hss\tenrm\hyperdef\hypernoname{page}\folio\folio\hss}}%
%

\def\draftmode{\message{ DRAFTMODE }\def\draftdate{{\rm preliminary draft:
\number\month/\number\day/\number\yearltd\ \ \hourmin}}%
\headline={\hfil\draftdate}\writelabels\baselineskip=20pt plus 2pt minus 2pt
 {\count255=\time\divide\count255 by 60 \xdef\hourmin{\number\count255}
  \multiply\count255 by-60\advance\count255 by\time
  \xdef\hourmin{\hourmin:\ifnum\count255<10 0\fi\the\count255}}}
\def\nolabels{\def\wrlabeL##1{}\def\eqlabeL##1{}\def\reflabeL##1{}}
\def\writelabels{\def\wrlabeL##1{\leavevmode\vadjust{\rlap{\smash%
{\line{{\escapechar=` \hfill\rlap{\sevenrm\hskip.03in\string##1}}}}}}}%
\def\eqlabeL##1{{\escapechar-1\rlap{\sevenrm\hskip.05in\string##1}}}%
\def\reflabeL##1{\noexpand\llap{\noexpand\sevenrm\string\string\string##1}}}
\nolabels
%
\global\newcount\secno \global\secno=0
\global\newcount\meqno \global\meqno=1
\def\s@csym{}
\def\newsec#1{\global\advance\secno by1%
{\toks0{#1}\message{(\the\secno. \the\toks0)}}%
\global\subsecno=0\eqnres@t\let\s@csym\secsym\xdef\secn@m{\the\secno}\noindent
{\bf\hyperdef\hypernoname{section}{\the\secno}{\the\secno.} #1}%
\writetoca{{\string\hyperref{}{section}{\the\secno}{\the\secno.}} {#1}}%
\par\nobreak\medskip\nobreak}
\def\eqnres@t{\xdef\secsym{\the\secno.}\global\meqno=1\bigbreak\bigskip}
\def\sequentialequations{\def\eqnres@t{\bigbreak}}\xdef\secsym{}
\global\newcount\subsecno \global\subsecno=0
\def\subsec#1{\global\advance\subsecno by1%
{\toks0{#1}\message{(\s@csym\the\subsecno. \the\toks0)}}%
\ifnum\lastpenalty>9000\else\bigbreak\fi
\noindent{\it\hyperdef\hypernoname{subsection}{\secn@m.\the\subsecno}%
{\secn@m.\the\subsecno.} #1}\writetoca{\string\quad
{\string\hyperref{}{subsection}{\secn@m.\the\subsecno}{\secn@m.\the\subsecno.}}
{#1}}\par\nobreak\medskip\nobreak}
\def\appendix#1#2{\global\meqno=1\global\subsecno=0\xdef\secsym{\hbox{#1.}}%
\bigbreak\bigskip\noindent{\bf Appendix \hyperdef\hypernoname{appendix}{#1}%
{#1.} #2}{\toks0{(#1. #2)}\message{\the\toks0}}%
\xdef\s@csym{#1.}\xdef\secn@m{#1}%
\writetoca{\string\hyperref{}{appendix}{#1}{Appendix {#1.}} {#2}}%
\par\nobreak\medskip\nobreak}
%
%
\def\checkm@de#1#2{\ifmmode{\def\f@rst##1{##1}\hyperdef\hypernoname{equation}%
{#1}{#2}}\else\hyperref{}{equation}{#1}{#2}\fi}
\def\eqnn#1{\DefWarn#1\xdef #1{(\noexpand\relax\noexpand\checkm@de%
{\s@csym\the\meqno}{\secsym\the\meqno})}%
\wrlabeL#1\writedef{#1\leftbracket#1}\global\advance\meqno by1}
\def\f@rst#1{\c@t#1a\em@ark}\def\c@t#1#2\em@ark{#1}
\def\eqna#1{\DefWarn#1\wrlabeL{#1$\{\}$}%
\xdef #1##1{(\noexpand\relax\noexpand\checkm@de%
{\s@csym\the\meqno\noexpand\f@rst{##1}}{\hbox{$\secsym\the\meqno##1$}})}
\writedef{#1\numbersign1\leftbracket#1{\numbersign1}}\global\advance\meqno by1}
\def\eqn#1#2{\DefWarn#1%
\xdef #1{(\noexpand\hyperref{}{equation}{\s@csym\the\meqno}%
{\secsym\the\meqno})}$$#2\eqno(\hyperdef\hypernoname{equation}%
{\s@csym\the\meqno}{\secsym\the\meqno})\eqlabeL#1$$%
\writedef{#1\leftbracket#1}\global\advance\meqno by1}
\def\xeqn{\expandafter\xe@n}\def\xe@n(#1){#1}
\def\xeqna#1{\expandafter\xe@n#1}
\def\eqns#1{(\e@ns #1{\hbox{}})}
\def\e@ns#1{\ifx\UNd@FiNeD#1\message{eqnlabel \string#1 is undefined.}%
\xdef#1{(?.?)}\fi{\let\hyperref=\relax\xdef\next{#1}}%
\ifx\next\em@rk\def\next{}\else%
\ifx\next#1\xeqn#1\else\def\n@xt{#1}\ifx\n@xt\next#1\else\xeqna#1\fi
\fi\let\next=\e@ns\fi\next}

\def\DefWarn#1{\ifx\UNd@FiNeD#1\else
\immediate\write16{*** WARNING: the label \string#1 is already defined ***}\fi}
%
\newskip\footskip\footskip14pt plus 1pt minus 1pt 
\def\footnotefont{\ninepoint}\def\f@t#1{\footnotefont #1\@foot}
\def\f@@t{\baselineskip\footskip\bgroup\footnotefont\aftergroup\@foot\let\next}
\setbox\strutbox=\hbox{\vrule height9.5pt depth4.5pt width0pt}
\global\newcount\ftno \global\ftno=0
\def\foot{\global\advance\ftno by1\def\foot@rg{\hyperref{}{footnote}%
{\the\ftno}{\the\ftno}\xdef\foot@rg{\noexpand\hyperdef\noexpand\hypernoname%
{footnote}{\the\ftno}{\the\ftno}}}\footnote{$^{\foot@rg}$}}
%
\newwrite\ftfile
\def\footend{\def\foot{\global\advance\ftno by1\chardef\wfile=\ftfile
\hyperref{}{footnote}{\the\ftno}{$^{\the\ftno}$}%
\ifnum\ftno=1\immediate\openout\ftfile=\jobname.fts\fi%
\immediate\write\ftfile{\noexpand\smallskip%
\noexpand\item{\noexpand\hyperdef\noexpand\hypernoname{footnote}
{\the\ftno}{f\the\ftno}:\ }\pctsign}\findarg}%
\def\footatend{\vfill\eject\immediate\closeout\ftfile{\parindent=20pt
\centerline{\bf Footnotes}\nobreak\bigskip\input \jobname.fts }}}
\def\footatend{}
%
%
\global\newcount\refno \global\refno=1
\newwrite\rfile
\def\ref{[\hyperref{}{reference}{\the\refno}{\the\refno}]\nref}
\def\nref#1{\DefWarn#1%
\xdef#1{[\noexpand\hyperref{}{reference}{\the\refno}{\the\refno}]}%
\writedef{#1\leftbracket#1}%
\ifnum\refno=1\immediate\openout\rfile=\jobname.refs\fi
\chardef\wfile=\rfile\immediate\write\rfile{\noexpand\item{[\noexpand\hyperdef%
\noexpand\hypernoname{reference}{\the\refno}{\the\refno}]\ }%
\reflabeL{#1\hskip.31in}\pctsign}\global\advance\refno by1\findarg}
\def\findarg#1#{\begingroup\obeylines\newlinechar=`\^^M\pass@rg}
{\obeylines\gdef\pass@rg#1{\writ@line\relax #1^^M\hbox{}^^M}%
\gdef\writ@line#1^^M{\expandafter\toks0\expandafter{\striprel@x #1}%
\edef\next{\the\toks0}\ifx\next\em@rk\let\next=\endgroup\else\ifx\next\empty%
\else\immediate\write\wfile{\the\toks0}\fi\let\next=\writ@line\fi\next\relax}}
\def\striprel@x#1{} \def\em@rk{\hbox{}}
\def\lref{\begingroup\obeylines\lr@f}
\def\lr@f#1#2{\DefWarn#1\gdef#1{\let#1=\UNd@FiNeD\ref#1{#2}}\endgroup\unskip}

\def\addref#1{\immediate\write\rfile{\noexpand\item{}#1}} 
\def\listrefs{\footatend\vfill\supereject\immediate\closeout\rfile\writestoppt
\baselineskip=\footskip\centerline{{\bf References}}\bigskip{\parindent=20pt%
\frenchspacing\escapechar=` \input \jobname.refs\vfill\eject}\nonfrenchspacing}
\def\startrefs#1{\immediate\openout\rfile=\jobname.refs\refno=#1}
\def\xref{\expandafter\xr@f}\def\xr@f[#1]{#1}
\def\refs#1{\count255=1[\r@fs #1{\hbox{}}]}
\def\r@fs#1{\ifx\UNd@FiNeD#1\message{reflabel \string#1 is undefined.}%
\nref#1{need to supply reference \string#1.}\fi%
\vphantom{\hphantom{#1}}{\let\hyperref=\relax\xdef\next{#1}}%
\ifx\next\em@rk\def\next{}%
\else\ifx\next#1\ifodd\count255\relax\xref#1\count255=0\fi%
\else#1\count255=1\fi\let\next=\r@fs\fi\next}
%

%
\newwrite\ffile\global\newcount\figno \global\figno=1
\def\fig{fig.~\hyperref{}{figure}{\the\figno}{\the\figno}\nfig}
\def\nfig#1{\DefWarn#1%
\xdef#1{fig.~\noexpand\hyperref{}{figure}{\the\figno}{\the\figno}}%
\writedef{#1\leftbracket fig.\noexpand~\xfig#1}%
\ifnum\figno=1\immediate\openout\ffile=\jobname.figs\fi\chardef\wfile=\ffile%
{\let\hyperref=\relax
\immediate\write\ffile{\noexpand\medskip\noexpand\item{Fig.\ %
\noexpand\hyperdef\noexpand\hypernoname{figure}{\the\figno}{\the\figno}. }
\reflabeL{#1\hskip.55in}\pctsign}}\global\advance\figno by1\findarg}
\def\listfigs{\vfill\eject\immediate\closeout\ffile{\parindent40pt
\baselineskip14pt\centerline{{\bf Figure Captions}}\nobreak\medskip
\escapechar=` \input \jobname.figs\vfill\eject}}
\def\xfig{\expandafter\xf@g}\def\xf@g fig.\penalty\@M\ {}
\def\figs#1{figs.~\f@gs #1{\hbox{}}}
\def\f@gs#1{{\let\hyperref=\relax\xdef\next{#1}}\ifx\next\em@rk\def\next{}\else
\ifx\next#1\xfig #1\else#1\fi\let\next=\f@gs\fi\next}
\def\figin{\epsfcheck\figin}\def\figins{\epsfcheck\figins}
\def\epsfcheck{\ifx\epsfbox\UNd@FiNeD
\message{(NO epsf.tex, FIGURES WILL BE IGNORED)}
\gdef\figin##1{\vskip2in}\gdef\figins##1{\hskip.5in}
\else\message{(FIGURES WILL BE INCLUDED)}%
\gdef\figin##1{##1}\gdef\figins##1{##1}\fi}
\def\DefWarn#1{}
\def\figinsert{\goodbreak\midinsert}
\def\ifig#1#2#3{\DefWarn#1\xdef#1{fig.~\noexpand\hyperref{}{figure}%
{\the\figno}{\the\figno}}\writedef{#1\leftbracket fig.\noexpand~\xfig#1}%
\figinsert\figin{\centerline{#3}}\medskip\centerline{\vbox{\baselineskip12pt
\advance\hsize by -1truein\noindent\wrlabeL{#1=#1}\footnotefont%
{\bf Fig.~\hyperdef\hypernoname{figure}{\the\figno}{\the\figno}:} #2}}
\bigskip\endinsert\global\advance\figno by1}
\newwrite\lfile
{\escapechar-1\xdef\pctsign{\string\%}\xdef\leftbracket{\string\{}
\xdef\rightbracket{\string\}}\xdef\numbersign{\string\#}}
\def\writedefs{\immediate\openout\lfile=\jobname.defs \def\writedef##1{%
{\let\hyperref=\relax\let\hyperdef=\relax\let\hypernoname=\relax
 \immediate\write\lfile{\string\def\string##1\rightbracket}}}}%
\def\writestop{\def\writestoppt{\immediate\write\lfile{\string\pageno
 \the\pageno\string\startrefs\leftbracket\the\refno\rightbracket
 \string\def\string\secsym\leftbracket\secsym\rightbracket
 \string\secno\the\secno\string\meqno\the\meqno}\immediate\closeout\lfile}}
\def\writestoppt{}\def\writedef#1{}
\def\seclab#1{\DefWarn#1%
\xdef #1{\noexpand\hyperref{}{section}{\the\secno}{\the\secno}}%
\writedef{#1\leftbracket#1}\wrlabeL{#1=#1}}
\def\subseclab#1{\DefWarn#1%
\xdef #1{\noexpand\hyperref{}{subsection}{\secn@m.\the\subsecno}%
{\secn@m.\the\subsecno}}\writedef{#1\leftbracket#1}\wrlabeL{#1=#1}}
\def\applab#1{\DefWarn#1%
\xdef #1{\noexpand\hyperref{}{appendix}{\secn@m}{\secn@m}}%
\writedef{#1\leftbracket#1}\wrlabeL{#1=#1}}
\newwrite\tfile \def\writetoca#1{}
\def\leaderfill{\leaders\hbox to 1em{\hss.\hss}\hfill}
\def\writetoc{\immediate\openout\tfile=\jobname.toc
   \def\writetoca##1{{\edef\next{\write\tfile{\noindent ##1
   \string\leaderfill {\string\hyperref{}{page}{\noexpand\number\pageno}%
                       {\noexpand\number\pageno}} \par}}\next}}}
\newread\ch@ckfile
\def\listtoc{\immediate\closeout\tfile\immediate\openin\ch@ckfile=\jobname.toc
\ifeof\ch@ckfile\message{no file \jobname.toc, no table of contents this pass}%
\else\closein\ch@ckfile\centerline{\bf Contents}\nobreak\medskip%
{\baselineskip=12pt\footnotefont\parskip=0pt\catcode`\@=11\input\jobname.toc
\catcode`\@=12\bigbreak\bigskip}\fi}
\catcode`\@=12 
%
\edef\tfontsize{\ifx\answ\bigans scaled\magstep3\else scaled\magstep4\fi}
\font\titlerm=cmr10 \tfontsize \font\titlerms=cmr7 \tfontsize
\font\titlermss=cmr5 \tfontsize \font\titlei=cmmi10 \tfontsize
\font\titleis=cmmi7 \tfontsize \font\titleiss=cmmi5 \tfontsize
\font\titlesy=cmsy10 \tfontsize \font\titlesys=cmsy7 \tfontsize
\font\titlesyss=cmsy5 \tfontsize \font\titleit=cmti10 \tfontsize
\skewchar\titlei='177 \skewchar\titleis='177 \skewchar\titleiss='177
\skewchar\titlesy='60 \skewchar\titlesys='60 \skewchar\titlesyss='60
\def\titlefont{\def\rm{\fam0\titlerm}
\textfont0=\titlerm \scriptfont0=\titlerms \scriptscriptfont0=\titlermss
\textfont1=\titlei \scriptfont1=\titleis \scriptscriptfont1=\titleiss
\textfont2=\titlesy \scriptfont2=\titlesys \scriptscriptfont2=\titlesyss
\textfont\itfam=\titleit \def\it{\fam\itfam\titleit}\rm}
 \ifx\answ\bigans\else scaled\magstep1\fi
\ifx\answ\bigans\def\abstractfont{\tenpoint}\else
\font\absit=cmti10 scaled \magstep1
\font\abssl=cmsl10 scaled \magstep1
\font\absrm=cmr10 scaled\magstep1 \font\absrms=cmr7 scaled\magstep1
\font\absrmss=cmr5 scaled\magstep1 \font\absi=cmmi10 scaled\magstep1
\font\absis=cmmi7 scaled\magstep1 \font\absiss=cmmi5 scaled\magstep1
\font\abssy=cmsy10 scaled\magstep1 \font\abssys=cmsy7 scaled\magstep1
\font\abssyss=cmsy5 scaled\magstep1 \font\absbf=cmbx10 scaled\magstep1
\skewchar\absi='177 \skewchar\absis='177 \skewchar\absiss='177
\skewchar\abssy='60 \skewchar\abssys='60 \skewchar\abssyss='60
\def\abstractfont{\def\rm{\fam0\absrm}
\textfont0=\absrm \scriptfont0=\absrms \scriptscriptfont0=\absrmss
\textfont1=\absi \scriptfont1=\absis \scriptscriptfont1=\absiss
\textfont2=\abssy \scriptfont2=\abssys \scriptscriptfont2=\abssyss
\textfont\itfam=\absit \def\it{\fam\itfam\absit}\def\footnotefont{\tenpoint}%
\textfont\slfam=\abssl \def\sl{\fam\slfam\abssl}%
\textfont\bffam=\absbf \def\bf
{\fam\bffam\absbf}\rm}\fi
\def\tenpoint{\def\rm{\fam0\tenrm}
\textfont0=\tenrm \scriptfont0=\sevenrm \scriptscriptfont0=\fiverm
\textfont1=\teni  \scriptfont1=\seveni  \scriptscriptfont1=\fivei
\textfont2=\tensy \scriptfont2=\sevensy \scriptscriptfont2=\fivesy
\textfont\itfam=\tenit \def\it{\fam\itfam\tenit}\def\footnotefont{\ninepoint}%
\textfont\bffam=\tenbf \def\bf{\fam\bffam\tenbf}\def\sl{\fam\slfam\tensl}\rm}
\font\ninerm=cmr9 \font\sixrm=cmr6 \font\ninei=cmmi9 \font\sixi=cmmi6
\font\ninesy=cmsy9 \font\sixsy=cmsy6 \font\ninebf=cmbx9
\font\nineit=cmti9 \font\ninesl=cmsl9 \skewchar\ninei='177
\skewchar\sixi='177 \skewchar\ninesy='60 \skewchar\sixsy='60
\def\ninepoint{\def\rm{\fam0\ninerm}
\textfont0=\ninerm \scriptfont0=\sixrm \scriptscriptfont0=\fiverm
\textfont1=\ninei \scriptfont1=\sixi \scriptscriptfont1=\fivei
\textfont2=\ninesy \scriptfont2=\sixsy \scriptscriptfont2=\fivesy
\textfont\itfam=\ninei \def\it{\fam\itfam\nineit}\def\sl{\fam\slfam\ninesl}%
\textfont\bffam=\ninebf \def\bf{\fam\bffam\ninebf}\rm}
%
%

\hyphenation{anom-aly anom-alies coun-ter-term coun-ter-terms}
\def\inv{^{\raise.15ex\hbox{${\scriptscriptstyle -}$}\kern-.05em 1}}

\def\Dsl{\,\raise.15ex\hbox{/}\mkern-13.5mu D} 
\def\dsl{\raise.15ex\hbox{/}\kern-.57em\partial}
\def\del{\partial}

\def\lspace{\ifx\answ\bigans{}\else\qquad\fi}
\def\lbspace{\ifx\answ\bigans{}\else\hskip-.2in\fi} 

\def\boxeqn#1{\vcenter{\vbox{\hrule\hbox{\vrule\kern3pt\vbox{\kern3pt
	\hbox{${\displaystyle #1}$}\kern3pt}\kern3pt\vrule}\hrule}}}
\def\mbox#1#2{\vcenter{\hrule \hbox{\vrule height#2in
		\kern#1in \vrule} \hrule}}  
%
 \def\CO{{\cal O}} 
   
   \def\CU{{\cal U}}

\def\vev#1{\langle #1 \rangle}

\def\darr#1{\raise1.5ex\hbox{$\leftrightarrow$}\mkern-16.5mu #1}

\def\roughly#1{\raise.3ex\hbox{$#1$\kern-.75em\lower1ex\hbox{$\sim$}}}

\input amssym

\input epsf

\def\IZ{\relax\ifmmode\mathchoice
{\hbox{\cmss Z\kern-.4em Z}}{\hbox{\cmss Z\kern-.4em Z}} {\lower.9pt\hbox{\cmsss Z\kern-.4em Z}}
{\lower1.2pt\hbox{\cmsss Z\kern-.4em Z}}\else{\cmss Z\kern-.4em Z}\fi}

\newif\ifdraft\draftfalse
\newif\ifinter\interfalse
\ifdraft\draftmode\else\interfalse\fi
\def\journal#1&#2(#3){\unskip, \sl #1\ \bf #2 \rm(19#3) }
\def\andjournal#1&#2(#3){\sl #1~\bf #2 \rm (19#3) }

\def\frac#1#2{{#1\over#2}}

\def\vev#1{\langle#1\rangle}

\def\inbar{\,\vrule height1.5ex width.4pt depth0pt}
\def\IC{\relax\hbox{$\inbar\kern-.3em{\rm C}$}}
\def\IR{\relax{\rm I\kern-.18em R}}
\def\IP{\relax{\rm I\kern-.18em P}}

%
%


%
\catcode`\@=11
\def\slash#1{\mathord{\mathpalette\c@ncel{#1}}}
\overfullrule=0pt

\def\MM{{\cal M}}

\def\SS{{\cal S}}

\def\underrel#1\over#2{\mathrel{\mathop{\kern\z@#1}\limits_{#2}}}

\catcode`\@=12

\def\A{{\alpha}}
\def\B{{\beta}}
\def\C{{\gamma}}
\def\D{{\delta}}

%

\def\vev#1{\left\langle #1 \right\rangle}

\def\exp{{\rm exp}}


\def\[{[}
\def\]{]}

\def\comment#1{ }

%
\def\draftnote#1{\ifdraft{\baselineskip2ex
                 \vbox{\kern1em\hrule\hbox{\vrule\kern1em\vbox{\kern1ex
                 \noindent \underbar{NOTE}: #1
             \vskip1ex}\kern1em\vrule}\hrule}}\fi}
\def\internote#1{\ifinter{\baselineskip2ex
                 \vbox{\kern1em\hrule\hbox{\vrule\kern1em\vbox{\kern1ex
                 \noindent \underbar{Internal Note}: #1
             \vskip1ex}\kern1em\vrule}\hrule}}\fi}

%
%



%
%
%
%

%

\def\inv{^{-1}}


\def\ra{{\rightarrow}}

\def\cN{{\cal N}}

\def\figin{\epsfcheck\figin}\def\figins{\epsfcheck\figins}
\def\epsfcheck{\ifx\epsfbox\UnDeFiNeD
\message{(NO epsf.tex, FIGURES WILL BE IGNORED)}
\gdef\figin##1{\vskip2in}\gdef\figins##1{\hskip.5in}
\else\message{(FIGURES WILL BE INCLUDED)}%
\gdef\figin##1{##1}\gdef\figins##1{##1}\fi}
\def\DefWarn#1{}
\def\figinsert{\goodbreak\midinsert}
\def\ifig#1#2#3{\DefWarn#1\xdef#1{fig.~\the\figno}
\writedef{#1\leftbracket fig.\noexpand~\the\figno} %
\figinsert\figin{\centerline{#3}}\medskip\centerline{\vbox{\baselineskip12pt
\advance\hsize by -1truein\noindent\footnotefont{\bf
Fig.~\the\figno:} #2}}
\bigskip\endinsert\global\advance\figno by1}

\def \la {\langle}
\def \ra {\rangle}

\def \pa {\partial}

\lref\VenezianoYB{
  G.~Veneziano,
  ``Construction of a crossing - symmetric, Regge behaved amplitude for linearly rising trajectories,''
Nuovo Cim.\ A {\bf 57}, 190 (1968).
}

\lref\SiversIG{
  D.~Sivers and J.~Yellin,
  ``Review of recent work on narrow resonance models,''
Rev.\ Mod.\ Phys.\  {\bf 43}, 125 (1971).
}

\lref\GerchkovitzGXX{
  E.~Gerchkovitz, J.~Gomis, N.~Ishtiaque, A.~Karasik, Z.~Komargodski and S.~S.~Pufu,
  ``Correlation functions of Coulomb branch operators,''
[arXiv:1602.05971 [hep-th]].
}

\lref\BrowerEA{
  R.~C.~Brower, J.~Polchinski, M.~J.~Strassler and C.~I.~Tan,
  ``The Pomeron and gauge/string duality,''
JHEP {\bf 0712}, 005 (2007).
[hep-th/0603115].
}

\lref\KutasovXB{
  D.~Kutasov,
  ``Geometry on the space of conformal field theories and contact terms,''
Phys.\ Lett.\ B {\bf 220}, 153 (1989).
}
\lref\RanganathanNB{
  K.~Ranganathan,
   ``Nearby CFTs in the operator formalism: the role of a connection,''
Nucl.\ Phys.\ B {\bf 408}, 180 (1993)
[hep-th/9210090].
}
\lref\RanganathanVJ{
  K.~Ranganathan, H.~Sonoda and B.~Zwiebach,
   ``Connections on the state space over conformal field theories,''
Nucl.\ Phys.\ B {\bf 414}, 405 (1994).
[hep-th/9304053].
}

\lref\mandelstam{
S.~Mandelstam, ``Dual-resonance models." Physics Reports 13.6 (1974): 259-353.
}

\lref\FreundHW{
  P.~G.~O.~Freund,
  ``Finite energy sum rules and bootstraps,''
Phys.\ Rev.\ Lett.\  {\bf 20}, 235 (1968).
}

\lref\FrishmanDQ{
  Y.~Frishman, A.~Schwimmer, T.~Banks and S.~Yankielowicz,
  ``The Axial Anomaly and the Bound State Spectrum in Confining Theories,''
Nucl.\ Phys.\ B {\bf 177}, 157 (1981).
}

\lref\MeyerJC{
  H.~B.~Meyer and M.~J.~Teper,
  ``Glueball Regge trajectories and the pomeron: a lattice study,''
Phys.\ Lett.\ B {\bf 605}, 344 (2005).
[hep-ph/0409183].
}

\lref\FortinNQ{
  J.~F.~Fortin, K.~Intriligator and A.~Stergiou,
  ``Current OPEs in superconformal theories,''
JHEP {\bf 1109}, 071 (2011)
[arXiv:1107.1721 [hep-th]].
}

\lref\CoonYW{
  D.~D.~Coon,
 ``Uniqueness of the veneziano representation,''
Phys.\ Lett.\ B {\bf 29}, 669 (1969).
}

\lref\FairlieAD{
  D.~B.~Fairlie and J.~Nuyts,
  ``A fresh look at generalized Veneziano amplitudes,''
Nucl.\ Phys.\ B {\bf 433}, 26 (1995)
[hep-th/9406043].
}

\lref\PonomarevJQK{
  D.~Ponomarev and A.~A.~Tseytlin,
  ``On quantum corrections in higher-spin theory in flat space,''
[arXiv:1603.06273 [hep-th]]
}

\lref\StromingerTalk{
  A.~Strominger, Talk at Strings 2014, Princeton.
}

\lref\CostaMG{
  M.~S.~Costa, J.~Penedones, D.~Poland and S.~Rychkov,
  ``Spinning conformal correlators,''
JHEP {\bf 1111}, 071 (2011)
[arXiv:1107.3554 [hep-th]].
}

\lref\CamanhoAPA{
  X.~O.~Camanho, J.~D.~Edelstein, J.~Maldacena and A.~Zhiboedov,
  ``Causality constraints on corrections to the graviton three-point coupling,''
JHEP {\bf 1602}, 020 (2016)
[arXiv:1407.5597 [hep-th]].
}

\lref\GrossKZA{
  D.~J.~Gross and P.~F.~Mende,
  ``The high-energy behavior of string scattering amplitudes,''
Phys.\ Lett.\ B {\bf 197}, 129 (1987).
}

\lref\KarlinerHD{
  M.~Karliner, I.~R.~Klebanov and L.~Susskind,
  ``Size and shape of strings,''
Int.\ J.\ Mod.\ Phys.\ A {\bf 3}, 1981 (1988).
}

\lref\Hitoshi{
http://hitoshi.berkeley.edu/221b/scattering3.pdf
}

\lref\Asnin{
  V.~Asnin,
  ``On metric geometry of conformal moduli spaces of four-dimensional superconformal theories,''
JHEP {\bf 1009}, 012 (2010)
[arXiv:0912.2529 [hep-th]].
}

\lref\SusskindAA{
  L.~Susskind,
  ``Strings, black holes and Lorentz contraction,''
Phys.\ Rev.\ D {\bf 49}, 6606 (1994)
[hep-th/9308139].
}

\lref\FriedanHI{
  D.~Friedan and A.~Konechny,
  ``Curvature formula for the space of 2-d conformal field theories,''
JHEP {\bf 1209}, 113 (2012)
[arXiv:1206.1749 [hep-th]].
}

\lref\WessYU{
  J.~Wess and B.~Zumino,
  ``Consequences of anomalous Ward identities,''
Phys.\ Lett.\ B {\bf 37}, 95 (1971).
}

\lref\AspinwallMN{
  P.~S.~Aspinwall,
  ``K3 surfaces and string duality,''
in {\it Differential Geometry Inspired by String Theory}, ed. by S.-T~Yau
[hep-th/9611137].
}

\lref\LercheUY{
  W.~Lerche, C.~Vafa and N.~P.~Warner,
  ``Chiral rings in ${\cal N}=2$ superconformal theories,''
Nucl.\ Phys.\ B {\bf 324}, 427 (1989).
}

\lref\Brower{
R.~C.~Brower and J.~Harte.
``Kinematic constraints for infinitely rising Regge trajectories,"
Physical Review 164.5 (1967): 1841.
}

\lref\PappadopuloJK{
  D.~Pappadopulo, S.~Rychkov, J.~Espin and R.~Rattazzi,
  ``OPE convergence in conformal field theory,''
Phys.\ Rev.\ D {\bf 86}, 105043 (2012)
[arXiv:1208.6449 [hep-th]].
}

\lref\LegRed{
Askey, Richard. "Orthogonal expansions with positive coefficients." Proceedings of the American Mathematical Society 16.6 (1965): 1191-1194.
}

\lref\CardyVR{
  J.~L.~Cardy,
  ``Continuously varying exponents and the value of the central charge,''
J.\ Phys.\ A {\bf 20}, L891 (1987).
}

\lref\AmatiWQ{
  D.~Amati, M.~Ciafaloni and G.~Veneziano,
  ``Superstring collisions at Planckian energies,''
Phys.\ Lett.\ B {\bf 197}, 81 (1987).
}

\lref\CecottiKZ{
  S.~Cecotti,
  ``${\cal N}=2$ Landau-Ginzburg versus Calabi-Yau sigma models: nonperturbative aspects,''
Int.\ J.\ Mod.\ Phys.\ A {\bf 6}, 1749 (1991).
}
\lref\FreedmanZZ{
  D.~Z.~Freedman and A.~Van Proeyen,
  {\it Supergravity}, Cambridge U. Press (2012).
}

\lref\FrishmanDQ{
  Y.~Frishman, A.~Schwimmer, T.~Banks and S.~Yankielowicz,
  ``The axial anomaly and the bound state spectrum in confining theories,''
Nucl.\ Phys.\ B {\bf 177}, 157 (1981).
}

\lref\DixonFJ{
  L.~J.~Dixon, V.~Kaplunovsky and J.~Louis,
  ``On effective field theories describing $(2,2)$ vacua of the heterotic string,''
Nucl.\ Phys.\ B {\bf 329}, 27 (1990).
}

\lref\CecottiME{
  S.~Cecotti and C.~Vafa,
  ``Topological antitopological fusion,''
Nucl.\ Phys.\ B {\bf 367}, 359 (1991).
}

\lref\GomisYAA{
  J.~Gomis, P.~S.~Hsin, Z.~Komargodski, A.~Schwimmer, N.~Seiberg and S.~Theisen,
   ``Anomalies, conformal manifolds, and spheres,''
JHEP {\bf 1603}, 022 (2016)
[arXiv:1509.08511 [hep-th]].
}
\lref\SeibergVC{
  N.~Seiberg,
  ``Naturalness versus supersymmetric nonrenormalization theorems,''
Phys.\ Lett.\ B {\bf 318}, 469 (1993)
[hep-ph/9309335].
}

\lref\deBoerSS{
  J.~de Boer, J.~Manschot, K.~Papadodimas and E.~Verlinde,
  ``The Chiral ring of AdS$_3$/CFT$_2$ and the attractor mechanism,''
JHEP {\bf 0903}, 030 (2009)
[arXiv:0809.0507 [hep-th]].
}
\lref\CecottiME{
  S.~Cecotti and C.~Vafa,
  ``Topological antitopological fusion,''
Nucl.\ Phys.\ B {\bf 367}, 359 (1991).
}
\lref\WessYU{
  J.~Wess and B.~Zumino,
 ``Consequences of anomalous Ward identities,''
Phys.\ Lett.\ B {\bf 37}, 95 (1971).
}

\lref\NarainJJ{
  K.~S.~Narain,
   ``New heterotic string theories in uncompactified dimensions $< 10$,''
Phys.\ Lett.\ B {\bf 169}, 41 (1986).
}

\lref\SeibergPF{
  N.~Seiberg,
   ``Observations on the moduli space of superconformal field theories,''
Nucl.\ Phys.\ B {\bf 303}, 286 (1988).
}
\lref\deBoerSS{
  J.~de Boer, J.~Manschot, K.~Papadodimas and E.~Verlinde,
 ``The chiral ring of AdS$_3$/CFT$_2$ and the attractor mechanism,''
JHEP {\bf 0903}, 030 (2009)
[arXiv:0809.0507 [hep-th]].
}
\lref\OsbornQU{
  H.~Osborn,
   ``${\cal N}=1$ superconformal symmetry in four-dimensional quantum field theory,''
Annals Phys.\  {\bf 272}, 243 (1999)
[hep-th/9808041].
}
\lref\ZamolodchikovGT{
  A.~B.~Zamolodchikov,
   ``Irreversibility of the flux of the renormalization group in a $2D$ field theory,''
JETP Lett.\  {\bf 43}, 730 (1986), [Pisma Zh.\ Eksp.\ Teor.\ Fiz.\  {\bf 43}, 565 (1986)].
}
\lref\SalamonQK{
 S.~M.~Salamon,`` Quaternionic K{\"a}hler manifolds,''
  Invent. \ Math.,\ 67, 143–171.
}

\lref\SalamonCO{
	S.~M.~Salamon,``On the cohomology of Kahler and hyper-K\"ahler manifolds,''
  Topology, 35(1), pp.137-155.
}

\lref\OGradyMT{
	K.~G.~O'Grady,``Desingularized moduli spaces of sheaves on a K3,''
 J. fur die reine und angew.
 Math., 1999, 512, pp. 49-117. Preprint version, arXiv:9708009v2 [math.AG].
}

 \lref\OGradyMS{
 	K.~G.~O'Grady,``A new six-dimensional irreducible symplectic variety,''
 	J. Algebraic Geom., 2003,
 	12, pp. 435-505.
 }

\lref\OGradyHD{
	K.~G.~O'Grady,``Higher-dimensional analogues of K3 surfaces,''
	Current developments in algebraic geometry 59 (2010): 257-293.
}

\lref\AlexeKS{
	D.~V.~Alekseevsky and S.~Marchiafava,
	`` A twistor construction of K{\"a}hler submanifolds of a quaternionic K{\"a}hler manifold,''
	Ann. Mat. Pura Appl. (4) 184 (2005), no. 1, 53-74.
}

\lref\HarveyALE{
	J.~A.~Harvey, S.~Lee and S.~Murthy,
 ``Elliptic genera of ALE and ALF manifolds from gauged linear sigma models,''
	JHEP {\bf 1502}, 110 (2015).
	[arXiv:1406.6342 [hep-th]].
}

\lref\Strominger{
  A.~Strominger,
  ``Special geometry,''
Commun.\ Math.\ Phys.\  {\bf 133}, 163 (1990).
}

\lref\ttstar{
  S.~Cecotti and C.~Vafa,
  ``Topological antitopological fusion,''
Nucl.\ Phys.\ B {\bf 367}, 359 (1991).
}

\lref\Candelas{
  P.~Candelas and X.~de la Ossa,
  ``Moduli space of Calabi-Yau manifolds,''
Nucl.\ Phys.\ B {\bf 355}, 455 (1991).
}

\lref\GreenQU{
	M.~B.~Green and N.~Seiberg,
	``Contact interactions in superstring theory,''
	Nucl.\ Phys.\ B {\bf 299}, 559 (1988).
}

\lref\AtickGY{
	J.~J.~Atick, L.~J.~Dixon and A.~Sen,
	``String calculation of Fayet-Iliopoulos D terms in arbitrary supersymmetric compactifications,''
	Nucl.\ Phys.\ B {\bf 292}, 109 (1987).
}

\lref\DineGJ{
	M.~Dine, I.~Ichinose and N.~Seiberg,
	``F Terms and D terms in string theory,''
	Nucl.\ Phys.\ B {\bf 293}, 253 (1987).
}

\lref\WittenBH{
	E.~Witten,
	``Superstring perturbation theory revisited,''
	[arXiv:1209.5461 [hep-th]].
}
	
\lref\DineSeiberg{
M.~Dine and N.~Seiberg,
``Microscopic knowledge from macroscopic physics in string theory,''
Nucl.\ Phys.\ B {\bf 301}, 357 (1988).
}

\lref\GreenSeiberg{
M.~B.~Green and N.~Seiberg,
 ``Contact interactions in superstring theory,''
Nucl.\ Phys.\ B {\bf 299}, 559 (1988).
}

\lref\WittenYC{
  E.~Witten,
  ``Phases of N=2 theories in two-dimensions,''
Nucl.\ Phys.\ B {\bf 403}, 159 (1993).
[hep-th/9301042].
}

\lref\GreenDA{
  D.~Green, Z.~Komargodski, N.~Seiberg, Y.~Tachikawa and B.~Wecht,
  ``Exactly Marginal Deformations and Global Symmetries,''
JHEP {\bf 1006}, 106 (2010).
[arXiv:1005.3546 [hep-th]].
}

\lref\DoroudXW{
	N.~Doroud, J.~Gomis, B.~Le Floch and S.~Lee,
	``Exact results in D=2 supersymmetric gauge theories,''
	JHEP {\bf 1305}, 093 (2013).
	[arXiv:1206.2606 [hep-th]].
}
\lref\GomisWY{
	J.~Gomis and S.~Lee,
	``Exact K\"ahler potential from gauge theory and mirror symmetry,''
	JHEP {\bf 1304}, 019 (2013).
	[arXiv:1210.6022 [hep-th]].
}
\lref\DoroudPKA{
	N.~Doroud and J.~Gomis,
	``Gauge theory dynamics and K\"ahler potential for Calabi-Yau complex moduli,''
	JHEP {\bf 1312}, 99 (2013).
	[arXiv:1309.2305 [hep-th]].
}
\lref\BeniniUI{
	F.~Benini and S.~Cremonesi,
 	``Partition functions of ${\cal N}=(2,2)$ gauge theories on S$^{2}$ and vortices,''
	Commun.\ Math.\ Phys.\  {\bf 334}, no. 3, 1483 (2015).
	[arXiv:1206.2356 [hep-th]].
}

\lref\WB{
  J.~Wess and J.~Bagger,
  ``Supersymmetry and supergravity,''
Princeton, USA: Univ. Pr. (1992) 259 p.
}

\lref\JockersDK{
	H.~Jockers, V.~Kumar, J.~M.~Lapan, D.~R.~Morrison and M.~Romo,
	``Two-sphere partition functions and Gromov-Witten invariants,''
	Commun.\ Math.\ Phys.\  {\bf 325}, 1139 (2014).
	[arXiv:1208.6244 [hep-th]].
}

\lref\GerchkovitzGTA{
	E.~Gerchkovitz, J.~Gomis and Z.~Komargodski,
	``Sphere partition functions and the Zamolodchikov metric,''
	JHEP {\bf 1411}, 001 (2014).
	[arXiv:1405.7271 [hep-th]].
}
\lref\GomisYAA{
	J.~Gomis, P.~S.~Hsin, Z.~Komargodski, A.~Schwimmer, N.~Seiberg and S.~Theisen,
	``Anomalies, conformal manifolds, and spheres,''
	JHEP {\bf 1603}, 022 (2016).
	[arXiv:1509.08511 [hep-th]].
}

\lref\KatzHT{
	S.~H.~Katz, D.~R.~Morrison and M.~R.~Plesser,
	 ``Enhanced gauge symmetry in type II string theory,''
	Nucl.\ Phys.\ B {\bf 477}, 105 (1996).
	[hep-th/9601108].
}

\lref\BershadskySP{
	M.~Bershadsky, C.~Vafa and V.~Sadov,
	 ``D strings on D manifolds,''
	Nucl.\ Phys.\ B {\bf 463}, 398 (1996).
	[hep-th/9510225].
}

\lref\DixonBG{
  L.~J.~Dixon,
  ``Some World Sheet Properties Of Superstring Compactifications, On Orbifolds And Otherwise.''
Proceedings of the 1987 ICTP Summer Workshop.
}

\lref\BershadskyCX{
	M.~Bershadsky, S.~Cecotti, H.~Ooguri and C.~Vafa,
	 ``Kodaira-Spencer theory of gravity and exact results for quantum string amplitudes,''
	Commun.\ Math.\ Phys.\  {\bf 165}, 311 (1994).
	[hep-th/9309140].
}

\lref\BaggioIOA{
	M.~Baggio, V.~Niarchos and K.~Papadodimas,
	 ``$tt^{*}$ equations, localization and exact chiral rings in 4d $\cN$ =2 SCFTs,''
	JHEP {\bf 1502}, 122 (2015).
	[arXiv:1409.4212 [hep-th]].
}

\lref\IntriligatorFF{
	K.~A.~Intriligator and W.~Skiba,
	 ``Bonus symmetry and the operator product expansion of N=4 SuperYang-Mills,''
	Nucl.\ Phys.\ B {\bf 559}, 165 (1999).
	[hep-th/9905020].
}

\lref\AliUT{
	A.~Ali,
	 ``Classification of two-dimensional N=4 superconformal symmetries,''
	[hep-th/9906096].
}

\lref\LinDSA{
  Y.~H.~Lin, S.~H.~Shao, Y.~Wang and X.~Yin,
  ``Supersymmetry Constraints and String Theory on K3,''
JHEP {\bf 1512}, 142 (2015).
[arXiv:1508.07305 [hep-th]].
}

\lref\AndrianopoliZH{
  L.~Andrianopoli, R.~D'Auria and S.~Ferrara,
  ``Supersymmetry reduction of N extended supergravities in four-dimensions,''
JHEP {\bf 0203}, 025 (2002).
[hep-th/0110277].
}

\lref\Lighthill{
M.~J.~Lighthill, {\it An Introduction to Fourier Analysis
and Generalized Functions}, Cambridge Monographs on Mechanics,
Cambridge Univ. Press (1958).}

\Title{
\vbox{\baselineskip -6pt
\hbox{CALT-TH-2016-031, IPMU 16-016, PUPT-2513}
}}
{\vbox{
\centerline{Shortening Anomalies in Supersymmetric Theories}
}}

\vskip .1in
\centerline{Jaume Gomis,$^1$
Zohar Komargodski,$^2$
Hirosi Ooguri,$^{3,4}$
Nathan Seiberg,$^5$ and Yifan Wang$^6$}

\vskip .3in

\centerline{$^1$ Perimeter Institute for Theoretical Physics, Waterloo, Ontario, N2L 2Y5, Canada}

\centerline{$^2$ Weizmann Institute of Science, Rehovot 76100, Israel}

\centerline{$^3$ Walter Burke Institute for Theoretical Physics,
Caltech, Pasadena, CA 91125, USA}

\centerline{$^4$ Kavli IPMU, WPI, University of Tokyo, Kashiwa, 77-8583, Japan}

\centerline{$^5$ School of Natural Sciences, Institute for Advanced Study, Princeton, NJ 08540, USA}

\centerline{$^6$ Joseph Henry Laboratories, Princeton University, Princeton, NJ 08544, USA}

\vskip .3in

\centerline{ {\it Dedicated to John Schwarz on his 75th birthday}}

\vskip .3in

\noindent
We present new anomalies in two-dimensional ${\cal N}=(2,2)$
superconformal theories.  They obstruct the shortening
conditions of chiral and twisted chiral multiplets at coincident points.  This implies that    marginal couplings cannot be promoted
to background superfields in short representations. Therefore, standard results that follow from ${\cal N}=(2,2)$ spurion analysis are invalidated.
These anomalies appear only if supersymmetry is enhanced beyond 
${\cal N}=(2,2)$. These anomalies
explain why the conformal manifolds of the K3 and $T^4$
sigma models are not K\"ahler and do not factorize into
chiral and twisted chiral moduli spaces and why there are
no ${\cal N}=(2,2)$ gauged linear sigma models that cover these conformal manifolds. We also present these results from the point of view of the Riemann curvature of conformal manifolds.

\vskip .5in \noindent

 \Date{}

\newsec{Introduction}

Conformal Field Theories (CFT's) often come in a continuous family labeled by their exactly marginal couplings.  This family, known as the conformal manifold $\cal M$, is endowed with a canonical metric, the Zamolodchikov metric~\ZamolodchikovGT. The Zamolodchikov metric is determined by the two-point functions of the exactly marginal operators. The general Riemannian structure of conformal manifolds was first discussed in~\refs{\SeibergPF,\KutasovXB}.

Such conformal manifolds appear in theories with extended symmetries, such as supersymmetry or current algebra ($e.g.$,\ the $c=1$ models),
and in certain large $N$ theories.\foot{Some general arguments about when such families may exist can be found in~\CardyVR.}  These
conformal manifolds also play an important role in the AdS/CFT correspondence and on string worldsheets. In the former case, the conformal manifold  of the boundary CFT maps to the space of vacua in
the bulk Anti-de Sitter space (AdS).
In the latter case, the conformal manifold of the worldsheet theory
maps to the space of solutions of the equations of motion in spacetime.

One of the fundamental properties of
the conformal manifold of two-dimensional  ~~~ ${\cal N}=(2,2)$ superconformal field theories (SCFT's) is that it factorizes locally\foot{There are
examples where the conformal manifold is modded out by a discrete symmetry,
which prevents it from being a product globally~\refs{\BershadskySP,\KatzHT}. We thank D.
Morrison for a useful discussion about this point.  } into  the product of two K\"ahler manifolds,
\eqn\factor{
{\cal M}_{c}\times {\cal M}_{tc}\,.}
Coordinates of ${\cal M}_c$ and   ${\cal M}_{tc}$ are  the coupling constants  of  the exactly marginal operators constructed from the dimension  $(\frac{1}{2}, \frac{1}{2})$ operators in the chiral and twisted chiral rings~\LercheUY\ of the SCFT.  When the Virasoro central
charge is $c=9$ and the SCFT is realized as a
non-linear sigma model with a Calabi-Yau threefold as its target space,
the factorization~\factor\ was proven in~\DixonFJ\ by combining string theory worldsheet SCFT and  target space arguments.\foot{For SCFT's realizing  Calabi-Yau  compactifications,  ${\cal M}_c$ and   ${\cal M}_{tc}$ parameterize the moduli space of complex structure and complexified K\"ahler class.}
For other early papers on the ${\cal N}=(2,2)$ conformal manifold
from the target space and the worldsheet view points,
see~{\refs{\DineSeiberg,\Strominger}\ and~\refs{\DixonBG,\DixonFJ,\ttstar,\BershadskyCX},  respectively.

More recently, in~\GomisYAA\ the Weyl anomaly on the conformal manifold  of ${\cal N}=(2,2)$ theories was used to rederive the factorization~\factor. The argument in~\GomisYAA ~applies to any  ${\cal N}=(2,2)$ SCFT,   with no restriction on the Virasoro central charge. An  {\it assumption} made in \GomisYAA\  was that the coupling constants  parameterizing   ${\cal M}_c$ and   ${\cal M}_{tc}$  could be promoted  to supersymmetric dimension $(0,0)$ background chiral and twisted chiral multiplets, in the spirit of \SeibergVC.
 Factorization of the conformal manifold \factor\ then followed from the classification of 
  anomalies of the partition function under super-Weyl transformations.
  Alternatively, one can also  easily provide an argument for factorization
  \factor\   in the spirit of \Asnin.

On the other hand, it is well-known that the conformal manifold of the  $n$-dimensional  torus $T^n$ SCFT is locally \NarainJJ,
\eqn\torus{
{O(n,n)\over O(n)\times O(n)}\,,}
while the conformal manifold of the K3 SCFT is locally \SeibergPF ,
\eqn\kthree{
{O(4,20)\over O(4)\times O(20)}\,.}
More generally, the conformal manifold of ${\cal N}=(4,4)$ SCFT's is locally of
the form \refs{\CecottiKZ} (see also~\deBoerSS),
\eqn\nfour{
{O(4,n)\over O(4)\times O(n)}\, ,}
for some $n$.\foot{For  a sigma model with a hyper-K\"ahler target space $M$,  $n=h^{1,1}(M)$.}
These examples
appear to be at odds with the factorization \factor\ of the conformal manifold of $\cN=(2,2)$ SCFT's proven in \GomisYAA. Indeed, with the exception of the $n=2$ case in \torus, these  conformal manifolds do  not factorize
 locally  into a product of K\"ahler manifolds. In fact, they are not even K\"ahler manifolds. This is   in spite of the fact that the $T^n$ SCFT with $n$ even and the other
SCFT's   enjoy ${\cal N}=(2,2)$ superconformal symmetry.

  Another way of presenting this puzzle is the following.  Normally, when
  a particular global symmetry of a theory implies some special properties,
  extending that symmetry does not ruin those properties.  Here, we see a counter-example to that.  Specifically, if we consider a theory with
  ${\cal N}=(4,4)$ supersymmetry and view it as a special case of a
  ${\cal N}=(2,2)$ theory, we might conclude that the conformal manifold
  should be K\"ahler and that it should factorize as in  \factor .
  This conclusion turns out to be wrong. Later we will discuss additional properties of ${\cal N}=(2,2)$ theories that do not hold true in  theories with extended supersymmetry.

In this note we resolve this tension.
In a generic ${\cal N}=(2,2)$ SCFT, the operator product expansion (OPE)
 between a chiral multiplet $\CO$ and a
 twisted  chiral  multiplet $\tilde \CO$ of scaling dimensions
 $( \frac{1}{2},  \frac{1}{2})$ does not have poles. On the other hand,
in all the counter-examples to  factorization mentioned  above, the OPE has a pole\foot{ Throughout this note we study the theories in Euclidean space. However, we use Lorentzian signature notation with coordinates $x^{\pm\pm}$ etc., which are complex conjugates of each other.  The reason for using this notation is that we use complex conjugation notation on the charged chiral objects like the odd coordinates $\theta$, as if we are in Lorentzian signature.}
\eqn\OPEsi{\eqalign{
 \CO |(x_1) \, \tilde \CO | (x_2) \ \sim \
 {1\over x_1^{--}-x_2^{--}}\, J_{++} (x_2)+\cdots\, ,}}
where the symbol $|$ on the left-hand side picks up the bottom
component of each multiplet.
The operator $J_{++}$ appearing on the right-hand side
must be both chiral and twisted chiral, and
with scaling dimensions $(1,0)$
and R-charges $(2,0)$.  $J_{++}$ is the bottom component of a special ${\cal N}=(2,2)$ short multiplet, which we shall denote by  ${\cal J}_{++}$. The existence of such a multiplet
implies that the R-symmetry  is enhanced
and that the SCFT enjoys enlarged superconformal symmetry  beyond
${\cal N}=(2,2)$ supersymmetry.
For example, in the K3 SCFT, the current $J_{++}$ enlarges the
R-symmetry from $U(1)$ to $SU(2)$.\foot{It is important that the current $J_{++}$ which enlarges the R-symmetry
corresponds to a normalizable state in the SCFT.}

Let us imagine that we deform the SCFT with  exactly marginal operators
\eqn\exactdefa{
\int d^2x\, d\theta^+ d\theta^- \, \lambda\,  \CO+\int d^2x\,  d\theta^+ d\bar\theta^- \, \tilde\lambda\, \tilde \CO
+ c.c\,}
A powerful idea~\SeibergVC\ is to promote $\lambda$ and $\tilde\lambda$ to background chiral and twisted chiral superfields. We will see that whenever the OPE between $\CO$ and $\tilde \CO$
is singular as in
\OPEsi, then the couplings cannot be promoted to such background superfields due to an anomaly! We can either promote the $\lambda$s to background chiral superfields or the $\tilde\lambda$s to background twisted chiral superfields, but we cannot do both simultaneously. To our knowledge, this type of obstruction has not been discussed before.

Before we explain how this comes about let us explain the physics of promoting couplings to background fields. The effective action as a function of the background fields satisfies the required symmetries when the operators do not have contact terms that spoil those symmetries.
For example, if a conserved current is not conserved at coincident points then the background effective action will not be gauge invariant. Another example is if in a CFT the operator equation $T_\mu^\mu=0$ does not hold at coincident points, then the background effective action fails to depend just on the conformal class of the background metric. Similarly, in supersymmetric theories, for the effective action to depend on $\lambda$ and $\tilde \lambda$ as chiral and twisted chiral superfields, the operators $\CO$ and $\tilde\CO$ have to obey their defining equations
\eqn\defeqs{\bar D_{\pm} \CO=0~,\qquad  \bar D_+\tilde\CO=D_-\tilde \CO=0}
not only at separated points (which holds by definition) but also at coincident points.
Loosely speaking, one could say that the equations~\defeqs\ have to be obeyed off-shell.
As with all anomalies, our shortening anomaly can be understood as the failure of the partition function   to be invariant under certain background  field transformations.   
The operator equations~\defeqs\ at coincident points would be a consequence of the partition function being invariant under certain background superfield transformations, akin to the standard way  conservation laws follow from the invariance of the partition function under background field transformations. When the partition function is not invariant under these background field transformations, we encounter an anomaly. This point of view is elaborated in appendix~A.

We  will show that the OPE \OPEsi\ induces contact terms of the form 
 \eqn\introd{\eqalign{
 &\bar D_+^{(1)} \CO(Z_1)\,\tilde \CO(Z_2)\sim  \delta^{(2)}( z_{1\bar 2})\, \theta^+_{12}\,{\cal J}_{++}(Z_2)\cr
& \bar D_+^{(2)}  \CO(Z_1) \, \tilde \CO(Z_2)\sim \delta^{(2)}(z_{1\bar 2})\, \theta_{12}^+\, {\cal J}_{++}(Z_1)\,,}}
and that it is impossible to tune both of these contact terms away (the notation will be explained in section 3.)
Therefore, the effective action does not depend on the background coupling constants as if they were chiral and twisted chiral superfields. Some of the background couplings have to sit in long multiplets. We call this phenomenon a {\it shortening anomaly}.\foot{We would like to emphasize that the operators $\CO$, $\tilde\CO$ remain short in the standard situation where the couplings are constant. Indeed, $\lambda,\tilde\lambda$ have vanishing beta functions, since the operators $\CO,\tilde\CO$ have no operators to combine with (see~\GreenDA\ and also earlier literature, e.g.~\DixonBG).
In particular,~\OPEsi\ does not induce a beta function. The shortening anomaly is in the background fields, not in the operators. The operator equations are modified only in nontrivial  configurations for the background superfields (see equation~(1.9)).} The discussion above is reminiscent of the clash between conservation of the vector and axial current in theories with fermions, where
by adjusting counter-terms either symmetry can be preserved, but not both simultaneously.

As is standard in such situations, where some operator equations are violated at coincident points,  when we turn on nontrivial backgrounds for $\lambda$ and $\tilde\lambda$ then the contact terms~\introd\ lead to nontrivial operator  equations.
Depending on which counter-terms we choose, one of the equations below has to be true:
\eqn\operatorialii{\eqalign{
&\bar D_+ \CO \sim \bar D_- \tilde \lambda\ {\cal J}_{++}\, ,\cr
&\bar D_+ \tilde \CO \sim D_-  \lambda\ {\cal J}_{++} \, .}}
For constant couplings, where only the bottom components of $\lambda$ and $\tilde \lambda$ are turned on, the operators $\CO$ and $\tilde\CO$ remain short, as they should.

The obstruction to promoting both chiral and twisted chiral couplings to short multiplets (which exists  only if supersymmetry is enhanced) invalidates the conclusions found in~\GomisYAA\ for such theories.\foot{The analysis in the spirit of~\Asnin\ is also invalidated by   this anomaly since an implicit assumption in such an analysis is  that it is possible to preserve the shortening conditions of all the dimension $(\frac{1}{2}, \frac{1}{2})$ chiral and twisted chiral operators both at separate and coincident points.}
 We therefore conclude that factorization \factor\ breaks  down only if the supersymmetry algebra is larger than ${\cal N}=(2,2)$.
The SCFT's with conformal manifolds \torus\ (with $n>2$ even), \kthree\ and \nfour\
are indeed endowed with such an operator of R-charge $(2,0)$ and have enhanced supersymmetries, thus resolving the paradox.\foot{Such an operator does not exist for the $T^2$ SCFT. And indeed, the $T^2$  conformal manifold \torus\ factorizes  into a product of K\"ahler manifolds, locally given by $SL(2)/U(1)\times SL(2)/U(1)$.}

This phenomenon is similar to a familiar situation in supergravity.  The target space of ${\cal N}=1$ supergravity in $4d$ is known to be K\"ahler (see~\WB\ for details).  But ${\cal N}=2$ supergravity with hypermultiplets has a quaternionic target space, which is not K\"ahler  (see~\FreedmanZZ\ for a review of the scalar manifolds of supergravities in various dimensions). There is no contradiction in that because the ${\cal N}=2$ theory is not merely a special ${\cal N}=1$ supergravity theory because it has a  multiplet including a graviphoton and a gravitino, which is not present in the generic $\cN=1$ theory.  The special ${\cal N}=(2,2)$ multiplet ${\cal J}_{++}$ that resolves our puzzle, which includes a conserved spin-${3\over 2}$  current,  plays a similar role to the graviphoton multiplet in this supergravity analog. For a related discussion see \AndrianopoliZH. 

We can relate our discussion to supergravity more directly if
we view our $2d$ models as worldsheet theories for strings and we study their spacetime description.  The low energy description of string compactifications on Calabi-Yau threefolds, described by $\cN=(2,2)$ SCFT's on the worldsheet, is captured by four-dimensional $\cN=2$ supergravity, whose scalar manifold indeed takes the factorized form~\factor. On the other hand, compactifications on manifolds leading to  enhanced spacetime supersymmetry, such as $T^6$ or K3 $\times T^2$, are instead described by  $\cN=8$ or $\cN=4$ supergravity. The scalar manifold in these theories does not factorize.\foot{The conformal manifold of the worldsheet SCFT is a subspace of the supergravity scalar manifold.}

As with all 't Hooft-like anomalies, our analysis leads to  theorems about  the non-existence of certain renormalization group (RG) flows. If there exists a strictly ${\cal N}=(2,2)$ supersymmetric RG flow in which all the infrared marginal couplings are realized as chiral or twisted chiral couplings along the flow, then it is guaranteed that the corresponding coupling constants in the infrared are in short representations and hence there is no shortening anomaly. Therefore, the conformal manifold would have to factorize into a chiral K\"ahler manifold and a twisted chiral K\"ahler manifold.
Therefore, one can immediately conclude that there {\it cannot} exist an ${\cal N}=(2,2)$ RG flow that  realizes the full conformal manifold of the K3 SCFT. Indeed,  constructions of gauged linear sigma models  (GLSM's)~\WittenYC\ which lead  to subspaces of the K3 conformal manifold are known (see for example~\refs{\AspinwallMN,\CecottiKZ}),
but it has never been possible to embed the K3 SCFT in a UV completion
 that covers the full  conformal manifold. The same holds for $T^4$ but not for  $T^2$, which does not have a shortening anomaly, and indeed admits an ${\cal N}=(2,2)$ GLSM representation that realizes the complete $T^2$ SCFT conformal manifold.\foot{The GLSM is an ${\cal N}=(2,2)$ $U(1)$ gauge theory with chiral
 multiplets $(X_1,X_2,X_3,P)$ of charges $(1,1,1,-3)$ with a superpotential
 $W=PG_3(X)$, where $G_3(X)$ is a homogeneous polynomial of degree 3. The twisted chiral coupling is realized by the complexified FI parameter and the  chiral coupling by the single complex parameter in $G_3(X)$.} We now see that the obstruction for K3 and $T^4$ is due to an anomaly in the infrared that must be matched in the ultraviolet.\foot{Our analysis however does not rule out ${\cal N}=(4,4)$ supersymmetric RG flows that would cover the full conformal manifold of sigma models on K3 or $T^4$.}

It is interesting to relate this discussion to recent developments concerning the $S^2$ partition functions of ${\cal N}=(2,2)$ SCFT's~\refs{\DoroudXW\BeniniUI\GomisWY-\DoroudPKA}. There are two inequivalent ways to compactify such theories on $S^2$ and they compute, respectively, the K\"ahler potentials of the chiral and twisted chiral deformations~\refs{\JockersDK,\GomisWY} (see also \refs{\GerchkovitzGTA,\GomisYAA}). However, this statement is not meaningful, if the total space does not factorize as in~\factor! Therefore, when our anomaly is present, also this result about general ${\cal N}=(2,2)$ SCFT's is invalidated. (One can understand it again as being due to the failure of spurion analysis.) Because of the anomaly, there is no ${\cal N}=(2,2)$ UV completion  that would cover the full conformal manifold, therefore, what one can extract from the sphere partition function is at best the K\"ahler potential on some K\"ahler submanifolds of the conformal manifold. An alternative sphere compactification that utilizes the extended supersymmetry may exist and it may probe the full conformal manifold.

We also provide
a complementary  perspective on the anomalies
by studying the Riemann curvature of the conformal manifold in ${\cal N}=(2,2)$ SCFT's, extending
the previous work \refs{\DixonFJ,\CecottiME,\deBoerSS}.   The study of the mixed chiral and twisted chiral exactly marginal curvature components  leads us  to establish   a factorization theorem: The conformal manifold of an ${\cal N}=(2,2)$ SCFT fails to factorize, if and only if the SCFT is endowed with a  conserved current of  R-charges $(2,0)$,
precisely the same operator $J_{++}$ responsible for the shortening anomaly.

The outline of the paper is as follows. In section 2 we set the stage by discussing the OPE of chiral and twisted chiral superfields.   In section~3 we show that if a particular short representation ${\cal J}_{++}$ appears in the OPE,
then one inevitably finds an obstruction to imposing the shortening conditions simultaneously on both the chiral and twisted chiral superfields.  In section~4 the same result is established  by analyzing  the Riemann curvature tensor of the   conformal manifold of ${\cal N}=(2,2)$ SCFT's. In section~5  we study the curvature of the conformal manifold of SCFT's endowed with the small ${\cal N}=(4,4)$ superconformal algebra and prove that the conformal manifold of such theories indeed takes the form~\nfour. In this section we further show that the extended supercharges acquire non-trivial holonomies when transported around the conformal manifold.

In Appendix A, we discuss the shortening anomaly from the Wess-Zumino point of view. Some of the calculations of the various curvatures are presented in Appendices B, C, and D.

 \newsec{OPE of Chiral and Twisted Chiral  Operators in ${\cal N}=(2,2)$ SCFT's}

 Around a given point on the conformal manifold of an   $\cN=(2,2)$ SCFT an exactly marginal operator can be realized as the top component of a chiral multiplet $\CO$ with  $U(1)_+\times U(1)_-$ R-charges $(1,1)$ or as the
  top component of a twisted chiral multiplet $\tilde \CO$ with R-charges $(1,-1)$.
  These multiplets obey the shortening conditions
 \eqn\BPScond{\eqalign{
 \bar D_+ \CO&=0\qquad \qquad  \bar D_- \CO=0\,,\cr
  \bar D_+ \tilde \CO&=0\qquad \qquad  D_- \tilde \CO=0\,.
 }}
 Our aim  is to determine to what extent these shortening conditions can be maintained as we explore the conformal manifold of an $\cN=(2,2)$ SCFT.

Monitoring  the shortening conditions \BPScond\ under an exactly marginal perturbation
\eqn\exactdef{
\int d^2x\, d\theta^+ d\theta^- \, \lambda\,  \CO+\int d^2x\,  d\theta^+ d\bar\theta^- \, \tilde\lambda\, \tilde \CO
+ c.c\,}
 leads us to analyze the contact terms in
 \eqn\ope{\eqalign{
 \left[ \bar D_{\pm} \CO(Z_1)\right]\, \tilde \CO(Z_2), \qquad
\CO(Z_1) \, \left[  \bar D_{+}\tilde \CO(Z_2)\right]  ,\qquad
  \CO(Z_1) \, \left[  D_{-} \tilde \CO(Z_2)\right] \,,
 }}
  where $Z_I\equiv(x_I^{\pm\pm},\theta^{\pm}_I,\bar \theta^{\pm}_I)$ are points in superspace. In this section we determine the operator product expansion (OPE) of a chiral and a twisted chiral multiplets, whose top components yield  marginal operators, leaving the analysis of   contact terms to the following section.

In supersymmetry, it is often useful to
employ  spurion analysis. E.g., the coupling constants of
chiral operators $\CO$ are promoted to
background chiral multiplets \SeibergVC . This procedure makes sense
only if the operator equation $ \bar D_{\pm} \CO=0$
is respected also at
coincident points (loosely speaking, we can say that $\CO$ is chiral
off-shell). This is because when we write the
partition function depending on some background fields, by taking
derivatives with respect to the background fields, we can probe the
correlation functions of the corresponding operators both at separated
and at coincident points. By definition, operators equations are always
obeyed at separated points, but they may fail at coincident points.
The famous (continuous) 't Hooft anomalies arise when a conservation
equation is not obeyed at coincident points. As a consequence, the
partition function depending on the  associated background fields
does not obey the naively expected equations  (in the famous case of
the chiral anomaly, the partition function is not gauge invariant).
 Similarly, we can couple twisted chiral background fields to
twisted chiral operators as long as the corresponding shortening
conditions are valid off-shell.

 A general method that guarantees that some conservation equation
is obeyed also at coincident points is to construct a regularization
obeying the conservation equation. This automatically
tunes the contact terms in the infrared to zero. Such a regularization
may not exist, if there is a genuine anomaly in the conservation equation.
In the next section we will establish the existence of some
contact terms that violate the  shortening
conditions \BPScond\ at coincident points. Therefore, the standard
arguments relying on the selection rules of background superfields
are not valid. This also means that, when this occurs,
 ${\cal N}=(2,2)$ RG flows that contain all the infrared chiral and
twisted chiral couplings cannot exist.

 We will show that such a subtlety in the spurion analysis takes place in
  an ${\cal N}=(2,2)$ SCFT in two dimensions
  when it has  an operator of dimensions
  $(1,0)$ and R-charges $(2,0)$.
  Unitarity implies that this operator, which we denote by $J_{++}$,
  is a conserved current, obeying $\partial_{--} J_{++}=0$. Since $J_{++}$ carries a non-vanishing R-charge, its existence
 implies that  the supersymmetry is enhanced.

A typical (but not the only) example of supersymmetry
 enhanced by such an extra R-current is
the small ${\cal N}=4$ superconformal symmetry with $SU(2)_R$
symmetry.
In an
 ${\cal N}=(4,4)$ SCFT,
chiral and twisted chiral operators are rotated into each other by
the extra R-current as
\eqn\pair{
  J_{++}(x_1) \ \overline{\tilde \CO}|  (x_2)
  \sim \frac{1}{x_1^{++}-x_2^{++}} \ \CO| (x_2) ,}
where $\CO |$ denotes the bottom component of the operator $\CO$.
 By taking OPE's of both sides of the equation with another
 twisted chiral operator $\tilde \CO(x_3)$ (or, in other words, using the symmetry of the OPE coefficients),
 we find
\eqn\OPEs{\eqalign{
 \CO|(x_2) \, \tilde \CO| (x_3) \ \sim \
 {1\over x_2^{--}-x_3^{--}}\, J_{++} (x_3)+\cdots\,,}}
where
$\cdots$ encode the contributions of non-chiral operators and
descendants. For simplicity, we are suppressing some coefficients that we will make more explicit later.
In particular, when the ${\cal N}=(4,4)$ SCFT is realized as
a non-linear sigma model whose target space is a hyper-K\"ahler manifold
$M$, there is a unique holomorphic 2-form $\Omega \in H^{2,0}(M)$,
and the extra R-current is expressed
as $J_{++} = \Omega_{ab} \psi_+^a \psi_+^b$.
Chiral and twisted chiral operators are expressed as,
\eqn\sigmamodel{
\CO| = k_{a \bar b}\ \psi_+^a \bar\psi_-^{\bar b}, ~~~~
\tilde \CO| = k_{a \bar b} g^{\bar b c} \Omega_{cd}\ \psi_+^a
\psi_-^d,}
where $k \in H^{1,1}(M)$, $g_{a\bar b}$ is the K\"ahler
metric on $M$, and $\psi_\pm$ are fermions. The OPE
\OPEs\ then follows from $\psi_-^{\bar b}(x_3)\psi_-^d(x_2) \sim
\frac{ g^{\bar b d}}{x_3^{--} - x_2^{--}} $.

 More generally, one can show that the existence of
 $J_{++}$ of dimensions $(1,0)$ and R-charges $(2,0)$
 alone is sufficient for the pairing \pair\
 and therefore the OPE \OPEs\ follows  without assuming
the small ${\cal N}=4$ superconformal
 symmetry.  Indeed, since $J_{++}$ does not commute with the   $U(1)_+\times U(1)_-$ R-symmetry in an ${\cal N}=(2,2)$ SCFT,  the
 R-symmetry group must be larger than just $U(1)_+\times U(1)_-$ and the dimension $(\frac{1}{2}, \frac{1}{2})$  chiral and twisted chiral operators must furnish a representation of it. Therefore, $J_{++}$ cannot act trivially and hence some pairing as in~\pair\ and~\OPEs\ must be present.

Unitarity and supersymmetry further imply that the extra R-current $J_{++}$
is  the bottom component of a very short multiplet, which we denote
by ${\cal J}_{++}$.   It obeys
\eqn\shortJa{
\bar D_+\,  {\cal J}_{++} =  \bar D_-\,   {\cal J}_{++}=  D_-\,   {\cal J}_{++}=0\,.}
 For convenience, we summarize the R-charges of various objects used in
 the following:
\eqn\chargeass{\matrix{
& &U(1)_+  &U(1)_- \cr
&\theta^+&   1 & 0    & \cr
&\theta^-&   0 & 1    &\cr
&D_+&       -1  &           0\cr
&D_-&          0  &          -1\cr
&\CO &         1 &                 1 \cr
&\tilde \CO &            1   &            -1 \cr
&{\cal J}_{++} &          2 &             0 \cr
&{\cal J}_{--} &          0  &           2
}}
\vskip-1pt

\smallskip

Our final goal is to  monitor the shortening conditions~\BPScond\ as we explore the conformal manifold while preserving ${\cal N}=(2,2)$ supersymmetry. This prompts us to   determine   the OPE of the supermultiplets $\CO$,   $\tilde \CO$ and ${\cal J}_{++}$ in ${\cal N}=(2,2)$ superspace, which makes supersymmetry manifest.   
Introducing coordinates $(x,\theta,\bar\theta)$ (suppressing the $+,-$ indices) we define
 \eqn\Ds{\eqalign{
 D= {\partial\over \partial \theta}-i \bar\theta \partial~, \qquad  \qquad \qquad
  \bar D= -{\partial\over \partial \bar \theta }+i  \theta \partial\,,
 }}
 which generate
 \eqn\susy{
 \{D,\bar D\}=2i \partial\,.
 }
The chiral coordinates $y\equiv x-i\theta\bar\theta$ and $\bar y\equiv x+i\theta\bar\theta$ obey
\eqn\annih{
\bar D\, y=0\qquad \qquad\qquad D\, \bar y=0\,}
and are chiral/anti-chiral respectively, which we denote  also as $c/\bar c$. In this language $\CO$ is $(c,c)$, $\tilde \CO$ is $(c,\bar c)$ and ${\cal J}_{++}$ is simultaneously $(c,c)$ and $(c,\bar c)$.

We want to determine the dependence of three-point correlators on the superspace
position of the operators. We start with the supertranslation Ward identities.
Given two points in superspace $(x_1,\theta_1,\bar\theta_1)$ and
$(x_2,\theta_2,\bar\theta_2)$, we can define two  independent  even linear invariants:\foot{
The canonical invariants are constructed from the Maurer Cartan one-form $g^{-1}dg=(dx-i \theta d\bar\theta+id\theta\bar\theta, d\theta,d\bar\theta)$, where $g=\exp{\left(-i(x P+\theta Q+\bar \theta \bar Q)\right)}$. This yields the supertranslation  invariants $(\Delta_{12}\equiv x_1-x_2-i\theta_2\bar\theta_1+i\theta_1\bar\theta_2,\, \theta_{12}\equiv\theta_1-\theta_2,\, \bar\theta_{12}\equiv\bar\theta_1-\bar\theta_2)$. Note that $z_{1\,\bar 2}=\Delta_{12}-i \theta_{12}\bar\theta_{12}$ and $z_{\bar 1\, 2}=\Delta_{12}+i \theta_{12}\bar\theta_{12}$.}
\eqn\invariantts{\eqalign{
z_{1\bar 2}&\equiv
x_1-x_2-i\theta_2\bar\theta_1+i\theta_1\bar\theta_2 - i\theta_{12}\bar\theta_{12}=
y_1 -\bar y_2 +2i \theta_1\bar \theta_2\cr
z_{\bar 1 2}&\equiv
 x_1-x_2+i\theta_1\bar\theta_1+i\theta_2\bar\theta_2-2i\theta_2\bar\theta_1=
\bar y_1 - y_2 -2i \theta_2\bar \theta_1\,.
}}
 Supertranslation invariance implies that correlators depend on the position of operators through  $z_{i\, \bar j}$ and $z_{\,\bar i\,  j}$. The coordinates we have defined are rather convenient. Indeed, if the $i$-th operator  is chiral, the correlators depend  on $z_{i\, \bar j}$ only, while if it is anti-chiral they depend   on $z_{\,\bar i\,  j}$.

Our  correlator of interest is:
 \eqn\threepoint{
 \langle {\cal O}(Z_1)\, \tilde {\cal O}(Z_2)\, \overline {\cal J}_{++}(Z_3)\rangle\,,}
 where $Z_I\equiv(x_I^{\pm\pm},\theta^{\pm}_I,\bar \theta^{\pm}_I)$.
The shortening conditions,
\eqn\bps{\eqalign{
\bar D_+ {\cal O}&=\bar D_- {\cal O}=0~,\cr
\bar D_+ \tilde {\cal O}&=D_- \tilde {\cal O}=0~,\cr
D_+\, \overline {\cal J}_{++}&=\bar D_-\, \overline {\cal J}_{++}= D_-\, \overline {\cal J}_{++}=0\,~,}}
together with  supertranslational and rotational invariance imply that the correlator \threepoint\ depends on $z^{++}_{1\, \bar 3}$, $z^{++}_{2\, \bar 3}$ and $z^{--}_{1\, \bar 2}$. A subtlety in SCFT's that must   be taken  into account  is the existence of a superconformal invariant $X(Z_1,Z_2,Z_3)$ constructed out of  three   points in superspace~\OsbornQU. Superconformal invariance and nilpotency of $X(Z_1,Z_2,Z_3)$ imply that the most general  correlator  consistent with superconformal invariance is given by
\eqn\finalc{
\langle {\cal O}(Z_1)\, \tilde {\cal O}(Z_2)\, \overline{\cal J}_{++}(Z_3)\rangle={C\over
z^{++}_{1\, \bar 3}\, z^{++}_{2\, \bar 3}\, z^{--}_{1\, \bar 2}}\left[1+ a\, X(Z_1,Z_2,Z_3)\right]\,.}
Imposing that ${\cal O}(Z_1)$ is chiral forces $a=0$, as $X(Z_1,Z_2,Z_3)$ depends on $\bar\theta_1$~\OsbornQU. We therefore conclude that the correlator at separated points is
\eqn\finalc{
\langle {\cal O}(Z_1) \,\tilde {\cal O}(Z_2) \,\overline {\cal J}_{++}(Z_3)\rangle={C\over
z^{++}_{1\, \bar 3}\, z^{++}_{2\, \bar 3}\, z^{--}_{1\, \bar 2}}\,.}
Moreover, using that
\eqn\twopoint{
\langle
{\cal J}_{++}(Z_2)\,\overline{\cal J}_{++}(Z_3)\rangle=\left({1\over z^{++}_{2\, \bar 3}}\right)^2
}
 we obtain our desired OPE:
 \eqn\OPE{
 \CO(Z_1)\, \tilde \CO(Z_2)\sim {C\over z^{--}_{1\, \bar 2}}\,{\cal J}_{++}(Z_2)\,.}

The superspace correlators we have constructed obey, by construction,  the   shortening conditions~\bps\ at separated points. Our next task is to study the shortening conditions  at coincident points   \ope\ and determine whether counter-terms can be adjusted  so that  the shortening conditions for chirals and twisted chirals~\BPScond\ can be both simultaneously obeyed.

\newsec{Supersymmetric Contact Terms and the Shortening Anomaly}

The  OPE~\OPE\ may lead to some contact terms~\ope.\foot{  A related discussion appeared in \GreenQU, where it clarified the need for a contact term, which had been found earlier in \refs{\AtickGY,\DineGJ}.  A more modern discussion of that problem appeared in \WittenBH.} To understand these we need to study the  superspace derivatives (the derivatives with respect to the second argument follow from these)
\eqn\deriva{
D_+^{(1)}\left({1\over z_{1\,\bar 2}^{--}}\right)\,,\qquad \qquad
\bar D_+^{(1)}\left({1\over z_{1\,\bar 2}^{--}}\right)\,.}
Our strategy to compute~\deriva\ is to extend the well-known formula, $\del_{++}\left({1\over x_{12}^{--}}\right)=\pi\delta^{(2)}(x_{12})$, $i.e.$,
\eqn\greensf{
\del_{++}\left({1\over x_{12}^{--}}\right)=\del_{++}\del_{--}\log(|x_{12}|^2)={1\over 4}\square\log(|x_{12}|^2)=\pi\delta^{(2)}(x_{12})\, ,
}
to Green's function in superspace.
This line of inquiry makes manifest an inherent ambiguity in defining the derivatives~\deriva\ in superspace. In order to  define these derivatives we  {\it need}  to specify the behavior of Green's function in superspace  for the left-movers, and different choices yield different answers. This ambiguity is a manifestation of the fact that the pole $1/z_{1\,\bar 2}^{--}$ in ${\cal N}=(2,2)$ superspace is too singular to  yield an unambiguous distribution.\foot{On
the other hand, the
pole $1/x^{--}$ on the 2-plane of $(x^{++}, x^{--})$
can be extended to an unambiguous distribution because $\int d^2 x
f(x^{++},x^{--})/x^{--}$ is well-defined for any smooth function $f$. The pole $1/ z_{1\,\bar 2}^{--}$ in superspace, however, is akin to $1/(x^{--})^2$
and cannot be extended to an unambiguous distribution. The distribution defined via $1/(x^{--})^2\equiv -\partial_{--}(1/x^{--})$ corresponds to a particular choice of regularization.}
In fact, we will find an ambiguity of $\theta_{12}^+ \bar{\theta}_{12}^+
\delta^{(2)}(z_{1\bar 2})$ in defining  $1/ z_{1 \,\bar 2}^{--}$. This is analogous
to the well-known
ambiguity in the distribution $1/x$ on the real line
$x \in R$ by the addition of $\delta(x)$.\foot{$1/x$ is characterized by the fact that $x^n \cdot 1/x = x^{n-1}$ for all
positive integers $n$. Therefore, any ambiguity in $1/x$ must be annihilated by
multiplication by $x^n$, and the only distribution with this property is
proportional to $\delta(x)$  (See, for example, theorem 9 in \Lighthill). Indeed,
 the principal value $P(1/x)$ and
$1/(x \pm i \epsilon)$ differ by distributions proportional to $\delta(x)$.}
As a result,  the derivatives~\deriva\ suffer from certain ambiguities. However, we will find that, whichever choice we make for
these ambiguities, we cannot remove the
contact terms for both chiral and twisted chiral operators simultaneously.

We can compute~\deriva\ starting from the Green's functions  in superspace
\eqn\propchiral{
D_+^{(1)}D_-^{(1)}\log (z_{1\,\bar 2}^{++}z_{1\, \bar 2}^{--})= -4\pi \delta^{(2)}(z_{1\,\bar 2})\,\bar\theta_{12}^+\,\bar\theta_{12}^-\,
}
or alternatively
\eqn\proptchiral{
\bar D_+^{(1)}D_-^{(1)}\log (z_{\bar 1\, 2}^{++}z_{1\, \bar 2}^{--})=4\pi  \delta^{(2)}(z_{1\,\bar 2})\,\theta_{12}^+\,\bar\theta_{12}^-\,.}
Physically, \propchiral\ and \proptchiral\ can be interpreted as the Schwinger-Dyson equations or the Ward identities for the shift symmetry of   a chiral multiplet and a twisted chiral multiplet respectively. But these are not  the only   choices.  Consider the following identity
 \eqn\derivamix{
 \left({1\over z_{1\,\bar 2}^{--}}\right)D_-^{(1)} z_{1\,\bar 2}^{--}=
D_-^{(1)}\left[ a\log (z_{1\,\bar 2}^{++}z_{1\, \bar 2}^{--})+b\log (z_{\bar 1\, 2}^{++}z_{1\, \bar 2}^{--})\right]~,
 }
which holds as long as $a+b=1$. Using     equations \propchiral\ and \proptchiral\ we arrive at\foot{ Note that equations \propchiral\ and \proptchiral\ only fix    $ D_+^{(1)}\left({1/ z_{1\,\bar 2}^{--}}\right)$ and $ \bar D_+^{(1)}\left({1/ z_{1\,\bar 2}^{--}}\right)$ up to terms proportional to $\bar \theta_{12}^-$. However such terms are forbidden by the condition that $\bar D_{-}^{(1)}$ annihilates $D_+^{(1)}\left({1/ z_{1\,\bar 2}^{--}}\right)$ and $ \bar D_+^{(1)}\left({1/ z_{1\,\bar 2}^{--}}\right)$.    }
\eqn\optionmix{\eqalign{
 D_+^{(1)}\left({1\over z_{1\,\bar 2}^{--}}\right)&=-2\pi i a\,  \delta^{(2)}( z_{1\bar 2})\, \bar\theta^+_{12}\,,\cr \cr
 \bar D_+^{(1)}\left({1\over z_{1\,\bar 2}^{--}}\right)&=2\pi i b\,   \delta^{(2)}( z_{1\bar 2}) \,\theta^+_{12}
 \,.}}
 These expressions obey the supersymmetry algebra for all $a+b=1$
 \eqn\susyalg{
 \{D_+,\bar D_+\}=2i \partial_{++}\,.}
 Amongst these, there are distinguished  canonical choices
\eqn\derivs{\matrix{
&\hbox{} & D_+^{(1)}\left({1\over z_{1\, \bar 2}^{--}}\right)  &\bar D_+^{(1)}\left({1\over z_{1\, \bar 2}^{--}}\right)\cr
&& &\cr
& \hbox{ Preserves chiral Ward identity } &   -2\pi & 0    & \cr
&& &\cr
&\hbox{ Preserves twisted chiral Ward identity }  &   0 & 2\pi    &\cr
&& &\cr
&\hbox{ Symmetric violation of Ward identities} &     -\pi    &     \pi
 \cr    \cr
 }}
We note that there is no choice of $a$ that preserves simultaneously the chiral and twisted chiral Ward identity.  Even though the derivatives are subject to some ambiguities, we now proceed
to  unambiguously establish our shortening anomaly.

 Anomalies arise when we have an operator that should satisfy an operator equation, such as \BPScond,  and   one finds that such an equation is only correct at separated points while at coincident points there are various contact terms.\foot{By turning on suitable background fields the ambiguities in the contact terms can be described as an ambiguity in adding local counter-terms constructed out of the operators in the theory and the background fields.  When such background fields are present, the problem with contact terms can be uplifted to problems at separated points.}
 Establishing the anomaly amounts to showing that these contact terms cannot be removed by redefining the scheme since  scheme redefinitions change the theory by  contact terms.

We consider ${\cal N}=(2,2)$ supersymmetric contact terms     since we assume that the conformal manifold can be explored while maintaining ${\cal N}=(2,2)$ supersymmetry. Therefore, the most general OPE
 between a chiral and a twisted chiral superfield in a theory with an enhanced R-symmetry, now  allowing for supersymmetric contact terms, is given by
 \eqn\contactOPE{\CO(Z_1)\tilde \CO(Z_2)\sim {C\over z_{1\bar 2}^{--} }{\cal J}_{++}(Z_2)+r\, \delta^{(2)}(z_{1\bar 2}) \,\bar\theta^+_{12}\, \theta^+_{12}\, {\cal J}_{++}(Z_2)\,,}
where $r$ is a scheme dependent constant, which can be shifted by changing the scheme.

In order to establish our shortening anomaly we must show that it is not possible to tune the contact term $r$ such that the shortening conditions for a chiral and a twisted chiral multiplet can be maintained simultaneously. Since the contact term in~\contactOPE\ is annihilated by $\bar D^{(1)}_-$ and $D^{(2)}_-$, these shortening equations are automatically preserved.  Acting with $\bar D_+^{(1)}$ and $\bar D_+^{(2)}$ we get using~\optionmix\foot{The formula
$
 \bar D_+^{(2)}{1\over z_{1\bar 2}^{--} }=2\pi i a\, \delta^{(2)}(z_{1\bar 2})\theta_{12}^+$
 follows from  \optionmix.}
 \eqn\eqnA{\eqalign{\bar D_+^{(1)} \CO(Z_1)\,\tilde \CO(Z_2)&\sim 2 \pi i C\, b\,   \delta^{(2)}( z_{1\bar 2}) \,\theta^+_{12}\,{\cal J}_{++}(Z_2)-r\, \delta^{(2)}(z_{1\bar 2})\,\theta^+_{12} \,{\cal J}_{++}(Z_2)\,,\cr
 \bar D_+^{(2)}  \CO(Z_1) \, \tilde \CO(Z_2)&\sim 2\pi iC\, a\,\delta^{(2)}(z_{1\bar 2})\,\theta_{12}^+\,{\cal J}_{++}(Z_1)+r\, \delta^{(2)}(z_{1\bar 2})\, \theta^+_{12}\, {\cal J}_{++}(Z_1)\,.}}
Preserving    the  chiral {\it and}     twisted chiral   shortening conditions   along the conformal manifold requires tuning the coefficient of the supersymmetric counter-term to obey
 \eqn\cancelAA{\eqalign{
 \hbox{chiral}:&  \qquad r=2\pi i C b~,\cr
  \hbox{twisted chiral}:&  \qquad r=-2\pi i C a~.}}
 However,  since $a+b=1$, it is impossible to solve both equations. This implies that we cannot simultaneously preserve the chiral and twisted chiral shortening conditions along the conformal manifold. By tuning the contact term, we can either preserve the chiral or the twisted chiral constraint, but not both. This is our shortening anomaly.

 The fact that we cannot preserve both constraints
 simultaneously is analogous to the situation in two dimensions with vector and
  axial anomalies \FrishmanDQ,
  where  contact terms   cannot
 remove both anomalies simultaneously.
 For a  complementary derivation of our shortening anomaly in the
cohomological approach based on the Wess-Zumino
consistency conditions, see Appendix A.

The contact terms we encountered lead to an operatorial violation  of the shortening equations upon deforming the theory by   background superfield sources for the exactly marginal operators (as in~\exactdef).
The shortening conditions  become
\eqn\operatorialiia{\eqalign{
&\bar D_+ \CO \sim \bar D_- \tilde \lambda\ {\cal J}_{++}~,\cr
&\bar D_+ \tilde \CO \sim D_-  \lambda\ {\cal J}_{++}\,.
~}}
 This implies that promoting the couplings of dimension $(\frac{1}{2}, \frac{1}{2})$ chiral and twisted chiral operators to full-fledged background chiral and twisted chiral superfields is impossible whenever the theory includes the operator $J_{++}$. We emphasize that the violations in \operatorialiia~depend on the fermionic components   of the multiplet to which the couplings have been promoted. 

\newsec{Curvature of Conformal Manifold and Factorization}

A CFT with exactly marginal operators comes equipped  with additional   structure beyond the metric on the conformal manifold ${\cal M}$. Operators in the CFT are  sections of
vector bundles over ${\cal M}$. A canonical example of this is the set of exactly marginal operators, which are sections of the tangent bundle $T{\cal M}$.
Transporting  operators in the vector bundle  along ${\cal M}$ leads to operator mixing, which is   governed by    a connection  on the vector bundle \refs{\KutasovXB,\RanganathanNB,\RanganathanVJ}.  The curvature of these connections
captures geometrical and topological data about the vector bundle of operators in the CFT.

The curvature of the tangent bundle $T{\cal M}$ can be used to prove theorems about the conformal manifold ${\cal M}$. The computation in~\DixonFJ\ of the curvature of   $T{\cal M}$ in two-dimensional ${\cal N}=(2,2)$ SCFT's with $c=9$,
obtained by  combining   the worldsheet   with spacetime arguments,    was used to prove that the conformal manifold of such SCFT's factorizes~\factor. The result in~\DixonFJ\ follows from the vanishing of the mixed
chiral and twisted chiral curvature components, which implies that the holonomy group is  the direct product of two commuting subgroups, which in turn implies factorization~\factor.
In this section we establish the following result about any ${\cal N}=(2,2)$ SCFT by investigating the curvature on its conformal manifold explicitly: the conformal manifold of an ${\cal N}=(2,2)$ SCFT does not factorize if and only if the SCFT has a  conserved current with R-charges $(2,0)$.

We now proceed to compute the mixed components of the Riemann tensor of ${\cal M}$ using CFT techniques, in the spirit of~\refs{\KutasovXB,\RanganathanVJ,\deBoerSS,\FriedanHI}. The
 tangent bundle $T{\cal M}$ comes equipped with a metric compatible connection, whose curvature we would like to find. The curvature of this connection is determined by a certain four-point function of the exactly marginal operators, which we denote by ${\cal U}_i$. The formula for the curvature can be written as~\FriedanHI\foot{Note that our definition of the curvature includes an overall normalization of $4$ compared to that of \FriedanHI. This is the convention in which the special geometry relation  takes the standard form. In addition, we define as usual $O(\infty)=\lim_{x\rightarrow\infty} x^{2\Delta_O}O(x)$ while keeping all the other insertions fixed.}
\eqn\form{R_{ijkl}=-RV\int {d^2y\over  \pi}\log(y^{++}  y^{--}) \vev{\CU_i(0)\,\CU_k(y)\,\CU_l(1)\,\CU_j(\infty)}_c~.}
 The symbol 'RV' stands for the prescription where we cut out small discs around the fixed operators and remove the power-law divergent terms.\foot{The energy-momentum tensor would appear in the OPE with a $1/y^2$ singularity. However, it has an angular dependence and thus this singularity vanishes upon integrating over the angles. We recall that dimension  $(1,1)$ operators do not appear in the OPE since a nonzero OPE coefficient would lead to a beta function.}
The subscript $c$ stands for the connected correlator, which is defined by subtracting the three possible disconnected contributions to the four-point function. As shown in~\RanganathanVJ,  an explicit  formula for the curvature can be given in terms of the CFT data: spectrum of operators and OPE coefficients.

Extending the  ${\cal N}=(2,2)$ superconformal Ward identities first introduced in \DixonFJ\ and further exploited in \deBoerSS\ we can show that the only non-trivial  mixed chiral and twisted chiral moduli four-point function that needs to be studied is
\eqn\fourpoint{
\vev{F_i(x)\bar F_{\bar j}(y) \tilde F_{\tilde k}(z)\overline{\tilde F}_{\overline{\tilde\ell}}(w)},}
where we have denoted the exactly marginal operators constructed from operators in the chiral and twisted chiral ring by
\eqn\define{
F_i={1\over 2}\int d\theta^+d\theta^- \CO_i~,\qquad\qquad \tilde F_{\tilde i}={1\over 2}\int d\theta^+d\bar \theta^- \tilde \CO_{\tilde i}~.
}
Using the superconformal Ward identities introduced in \DixonFJ,
 the four-point function of   interest~\fourpoint\ can be expressed in terms of the   four-point function of chiral and twisted chiral operators of dimension $(\frac{1}{2}, \frac{1}{2})$
\eqn\wardd{\eqalign{ &\vev{F_i(x)\bar F_{\bar j}(y) \tilde F_{\tilde k}(z)\overline{\tilde F}_{\overline{\tilde\ell}}(w)}\cr
&
=\del_{y^{++}}\del_{y^{--}}\del_{z^{--}}\del_{w^{++}}\left[ {y^{++}-w^{++}\over z^{++}-x^{++} } {y^{--}-  z^{--}\over x^{--}- w^{--}}\vev{\CO_i(x)\bar \CO_{\bar j}(y)\tilde \CO_{ {\tilde k}}(z)\overline{\tilde \CO}_{\overline{\tilde\ell}}(w)}  \right]\,.}}

 We are now ready to compute the curvature using~\form. It follows from~\wardd\ that  the independent components of the Riemann curvature tensor are $R_{i{ {\tilde k}} {\overline{\tilde\ell}} \bar j}$ and $R_{i {\overline{\tilde\ell}} {\tilde k}\bar j}$. The connected component prescription in~\form\ can be extended to both sides of  \wardd, and therefore, by pulling the operator located at $x$ to infinity we arrive at
\eqn\twistedidentity{\eqalign{
&\vev{F_i(\infty)\bar F_{\bar j}(y) \tilde F_{\tilde k}(z)\overline{\tilde F}_{\overline{\tilde\ell}}(w)}_c\cr
&= \del_{y^{++}}\del_{y^{--}}\del_{z^{--}}\del_{w^{++}}\left[ (y^{++}-w^{++})(y^{--}-  z^{--})\vev{\CO_i(\infty)\bar \CO_{\bar j}(y)\tilde \CO_{\tilde k}(z)\overline{\tilde \CO}_{\overline{\tilde\ell}}(w)}_c  \right]\,.}}
Integrating by parts in~\form\ and remembering that we are integrating over the complex plane with disks  around the punctures removed, the answer reduces to contour integrals around the punctures.  In order to get a non-zero contribution to the curvature, the function
 \eqn\fourpointloc{
 g(y)\equiv  \del_{z^{--}}\del_{w^{++}}\left[ (y^{++}-w^{++})(y^{--}-  z^{--})\vev{\CO_i(\infty)\bar \CO_{\bar j}(y)\tilde \CO_{\tilde k}(z)\overline{\tilde \CO}_{\overline{\tilde\ell}}(w)}_c  \right]\biggr|_{z=0,w=1}
}
must either have a constant piece at $y=0$ or have a simple pole at $y=1$.\foot{More singular terms are removed by the prescription in \form. One can verify that there is no contribution from infinity.}
Let us analyze $g(y)$ near  $y=1$. This leads us to consider the OPE studied in section 2
\eqn\opebefo{
\bar \CO_{\bar j}(y)\overline{\tilde \CO}_{\overline{\tilde\ell}}(w)\sim {C_{\bar j{\overline{\tilde\ell}}}\over  y^{--}- w^{--}} \bar J_{++}(w)\,,}
where $\bar J_{++}$   has R-charges $(-2,0)$. Using 
\eqn\threepoint{
\vev{\bar J_{++}(w)\,\tilde \CO_{\tilde k}(z)\,\CO_i(\infty)}={C_{i \tilde k}\over w^{++}- z^{++}}~,}
we find that indeed near $y=1$ the function $g(y)$  has a simple pole
\eqn\polebeh{
 g(y)\sim {y^{++}\over  y^{--}-1}\,.}
A very similar analysis of the behavior of the function near $y=0$, where now the relevant OPE is
\eqn\openew{
 \CO_{\bar j}(y)\tilde \CO_{\tilde k}(z)\sim {C_{\bar j {\tilde k}}\over y^{++}-z^{++}}\bar J_{--}(z)~,\,}
demonstrates that there is no contribution to the curvature from the contour integral around $y=0$.

In conclusion, we have shown that there is a non-trivial   component of the mixed curvature tensor  between chiral and twisted chiral moduli   in an $\cN=(2,2)$ SCFT if and  only if the  SCFT
has a  current with R-charges $(2,0)$. In such a SCFT,  the Riemann curvature  is given by
\eqn\mixedcomps{R_{i{\tilde k}{\overline{\tilde\ell}} \bar j} \sim C_{\bar j {\overline{\tilde\ell}}}C_{i \tilde k} ~,\qquad
R_{i{\overline{\tilde\ell}} {\tilde k} \bar j} \sim C_{\bar j {\tilde k}}C_{i {\overline{\tilde\ell}}}\,,}
and    the conformal manifold ${\cal M}$ no longer factorizes.

\newsec{$\cN=(4,4)$ Conformal Manifolds}

The small ${\cal N}=4$ superconformal algebra, which has an $SU(2)_R$ R-symmetry 
(see e.g. \deBoerSS), 
is an important example of extended supersymmetry.
In this section, we use the formula \form\ to compute the curvature on the conformal manifold 
 of ${\cal N}=(4,4)$ theories and give a purely field-theoretic
derivation that the local
geometry of the coset is~\nfour.
We also study the bundle of ${\cal N}=4$ supercurrents over
the conformal manifold
and show that there is no consistent choice of an ${\cal N}=2$
subalgebra even over a local coordinate patch. This gives a geometric
perspective on our  shortening anomaly.

Let us introduce some notation first:
the left-moving $\cN=4$ supercurrents are denoted by $S_{+++}^{\A A}$, and
the $SU(2)_R$ currents by $J^{(\A\B)}_{++}$,
where $\A$ and $A$ are doublet indices for the R-symmetry $SU(2)_R$ and the outer automorphism $SU(2)_{{\rm out}}$
of the $\cN=4$ superconformal algebra, respectively.
 
We use the convention
$\epsilon^{12}=\epsilon_{21}=1$ for the invariant tensors $\epsilon_{\A\B},\epsilon_{AB}$ and their inverses.
For the right moving sector, we use dotted indices.
We   denote the weight $\left({1\over2},{1\over 2}\right)$
BPS primaries by $\CO_{i \A\dot \A}$, where $i=1,\dots,n$.
Their weight $(1,1)$ descendants
\eqn\ffdescendants{
F_i^{A\dot A}={1\over 8}\{Q^{\A A}_+ , [Q^{\dot \A \dot A}_- ,\CO_{a\A\dot \A}]\}
}
are exactly marginal operators which preserve the ${\cal N}=(4,4)$ superconformal symmetry  \refs{\SeibergPF,\DixonFJ,\deBoerSS}. These operators span the conformal manifold $\MM$ of ${\cal N}=(4,4)$ SCFT's.
Their two-point functions are
\eqn\ffZamob{
\la  \CO_{i\A\dot \A}(x)  \CO_{j\B\dot \B}(y) \ra
={ \eta_{ij} \epsilon_{\A\B}\epsilon_{\dot \A\dot\B} \over (x-y)^2  }
~,}
and
\eqn\ffZamot{
\la   F_{i A\dot A} (x)  F_{ j B\dot B}(y) \ra
={ \eta_{ij}\epsilon_{AB}\epsilon_{\dot A\dot B} \over (x-y)^4 }~,
}
where $\eta_{ij}$ is the Zamolodchikov metric. 

\subsec{Riemannian Curvature}

In the ${\cal N}=4$ notation above,
the formula \form\ for the Riemann curvature is expressed as
\eqn\RCintegral{
R_{i A \dot A;j B\dot B;k C\dot C;\ell D\dot D}=-RV \int {d^2 y\over  \pi} \log |y^2| \vev{F_{i A\dot A}(0)\,F_{j B\dot B} (y)\,  F_{k C\dot C} (1)\,F_{\ell D\dot D} (\infty )}_c.
}
By choosing an $\cN=2$ subalgebra generated by $S_{+++}^{11}$ and $S_{+++}^{22}$,
marginal operators for (anti-)chiral multiples are $F_{i1\dot 1}$ ($F_{i2\dot 2}$),
while those for (anti-)twisted chiral multiples are $F_{i1\dot 2}$ ($F_{i2\dot 1}$).

As shown in the previous
section, there are non-zero curvature components in mixed  chiral
and twisted chiral directions, which in the $\cN=4$ notation can be expressed as
\eqn\mixeda{
R_{i1\dot 1;j1\dot 2;k2\dot 1;\ell 2\dot 2}=R_{i1\dot 1;j2\dot 1;k1\dot 2;\ell 2\dot 2}=-{1\over   k}\eta_{ij}\eta_{k\ell },
}
where $k$ in the normalization factor is related to the Virasoro central
charge by $c=6k$.
By using the Bianchi identity, we can also determine\foot{The Bianchi identity $R_{i[jk\ell]}=0$ follows from the crossing symmetry of the four-point function \FriedanHI.}
\eqn\mixedb{
R_{i2\dot 1;j1\dot 2;k1 \dot 1;\ell 2\dot 2}=-{1\over   k}(\eta_{i\ell }\eta_{jk}-\eta_{i k}\eta_{j\ell}).
}
On the other hand, the curvatures in the purely chiral directions
are controlled by  the
special geometry relation for $c=9$ and its generalization for other $c$
 \refs{\Strominger,\ttstar,\BershadskyCX,\BaggioIOA},
 which we re-derive in Appendix B using the formula \form.
 In the $\cN=4$ notation, it takes the form
\eqn\SGf{
	R_{i1\dot 1;j2\dot 2;k1\dot 1;\ell 2\dot 2}=\eta_{ij}\eta_{k\ell}+\eta_{i\ell}\eta_{jk}-C_{ik}{}^IC_{j\ell}{}^J g_{IJ},
	}
where $C_{ik}^J$ are the chiral ring coefficients.

In Appendix D, we use the ${\cal N}=(4,4)$ superconformal symmetry to find a relation between the curvature
components \mixeda, \mixedb, and \SGf\ 
\eqn\curvID{
R_{i1\dot 1;j2\dot 2;k1\dot 1;\ell 2\dot 2}+R_{i2\dot 1;j 12;k1\dot 1;\ell 2\dot 2}+R_{i2\dot 1;j2\dot 2;k1\dot 1;\ell 1\dot 2}=0~.
}
This allows us to determine
\eqn\nonmix{
R_{i1\dot 1;j2\dot 2;k1\dot 1;\ell 2\dot 2}={1\over   k}( \eta_{i\ell }\eta_{jk}-\eta_{i k}\eta_{j\ell}+\eta_{ij}\eta_{k\ell})~.
}
Comparing this with \SGf,
we obtain as a by-product the following constraint on the chiral ring of any $\cN=(4,4)$ SCFT
 \eqn\chiralringID{
		C_{i k}^IC_{j\ell}^J g_{IJ}=\left(
		1-{1\over   k}
		\right)(\eta_{ij}\eta_{k\ell}+\eta_{i\ell}\eta_{kj})
		+{1\over   k} \eta_{ik}\eta_{j\ell}.
}
For an $\cN=(4,4)$ SCFT realized by a sigma model on a hyper-K\"ahler manifold $M$, it would be nice to verify that the cohomology ring of $M$ satisfies the constraint~\chiralringID.\foot{A curious observation is that~\chiralringID\ implies that there is a uniform bound on chiral ring coefficients (squared) associated with the R-charge $(1,1)$ chiral primaries. The bound takes the schematic form $C^2<2-1/k$. Note that this is very different from the ${\cal N}=(2,2)$ case, where the chiral ring coefficients can blow up, say, as at the conifold point.} Combining these results,  the full Riemannian curvature is 
	\eqn\riemannc{\eqalign{
			&R_{iA\dot A;jB\dot B;kC\dot C;\ell D\dot D}=
			-{1\over  k}( \eta_{ik}\eta_{j\ell}-\eta_{i\ell }\eta_{jk})
			\epsilon_{AB}\epsilon_{CD}  \epsilon_{\dot A\dot B}\epsilon_{\dot C\dot D}
			\cr
			&~~~~~~~~~~~-{1\over 2k}
			\eta_{ij}\eta_{k\ell}\left(
			\left(\epsilon_{\dot A \dot C}\epsilon_{\dot B \dot D}+\epsilon_{\dot A\dot D}\epsilon_{\dot B \dot C}\right)\epsilon_{  A B}\epsilon_{  C  D}
			 +\left(\epsilon_{AC}\epsilon_{BD}+\epsilon_{AD}\epsilon_{BC}\right)\epsilon_{\dot A\dot B}\epsilon_{\dot C\dot D}
			\right).
		}}
		This   implies that the conformal manifold of an ${\cal N}=(4,4)$ SCFT is locally the coset manifold~\nfour 
\eqn\ffcm
{
\MM={O(4,n)\over O(4)\times O(n)}\,.}

\subsec{Supercurrent Bundle }

Let us turn to the curvature
of the bundle of the $\cN=4$ supercurrents $S^{\A  A}_{+++}$,\foot{The $k$ dependent normalization is due to the  two-point function $\la S_{+++}^{\A A}(0) S_{+++}^{\B B}(\infty)\ra=4k \epsilon^{\A\B} \epsilon^{AB}$.}
\eqn\scintegral{
R_{iC\dot C j D\dot D   \A A \B  B }=-{1\over 4k}RV\int {d^2y\over \pi}\log(y^{++}  y^{--}) \vev{S_{\A A+++}(0)\,F_{iC\dot C} (y)\,F_{j D\dot D}(1)  S_{ \B B +++}(\infty)}_c .
}
In Appendix C, we determine the connected four-point function using $
{\cal N}=(4,4)$ Ward identities to be
\eqn\scfp{\eqalign{
\la
&S^{\A A}_{+++}(0)\,F^{C\dot C}_i(y)\,F^{D\dot D}_j(1)  S^{ \B B}_{+++}(\infty)
\ra_c
=
-2\epsilon^{\A\B}\epsilon^{\dot C\dot D} \eta_{ij}
\pa_{y^{++}}  \pa_{y^{--}} \left(
{\epsilon^{AC} \epsilon^{BD}\over  y (\bar y-1) }+{ \epsilon^{AD} \epsilon^{BC} y\over \bar y-1}
\right)~.
}}
The $y$-integral in \scintegral\ can then be performed 
\eqn\SCCL{R_{iC\dot C ;j D\dot D;   \A A ;\B  B }= -{1\over 2k}\eta_{ij}\epsilon_{\dot C\dot D}\epsilon_{\A\B}(\epsilon_{AC} \epsilon_{BD}+\epsilon_{AD} \epsilon_{BC})
~,}
and
\eqn\SCCR{R_{iC\dot C ;j D\dot D;   \dot \A  \dot A ;\dot\B  \dot B }= -{1\over 2k}\eta_{ij}\epsilon_{ C D}\epsilon_{\dot\A\dot\B}(\epsilon_{\dot A\dot C} \epsilon_{\dot B\dot D}+\epsilon_{\dot A \dot D} \epsilon_{\dot B\dot C})~.
}
The nontrivial $SU(2)_{{\rm out}}\times SU(2)_{{\rm out}}$ holonomies
shown in these curvatures means that
it is not possible to choose an $\cN=2$ subalgebra consistently,
even on a local patch of $\MM$.

Since the tangent bundle $T\MM$ is
a tensor product of the left and right supercurrent bundles and
the bundle of weight $\left({1\over 2},{1\over 2}\right)$ chiral primaries,
the curvature tensor for the supercurrents computed here
can be combined with the curvatures for
the chiral primaries computed
in \refs{\deBoerSS} to reproduce the Riemann
curvature \riemannc~on $\MM$.

Although $\MM$ is not K\"ahler, 
it has K\"ahler sub-manifolds. In fact, the maximal K\"ahler
sub-manifold of a quaternionic-K\"ahler manifold is middle-dimensional \refs{\AlexeKS}.
In our case, the maximal K\"ahler submanifold of $\MM$ is locally
\eqn\kahlersm{
\SS={O(2,n)\over O(2)\times O(n)}~.
}
If we only turn on the marginal couplings associated to chiral multiplets and explore that 
sub-manifold of $\MM$, then 
there is no shortening anomaly, and the argument of \GomisYAA\ leading to K\"ahlerity applies.
In $\cN=(4,4)$ SCFTs,  the subspace spanned by the marginal couplings associated to chiral or twisted chiral operators  corresponds to the sub-manifold~\kahlersm .

\bigskip
\centerline{\bf Acknowledgments}

We thank Kevin Costello, David Morrison, Kyriakos Papadodimas, Ronen Plesser, Adam Schwimmer, Stefan Theisen, and Edward Witten for useful discussions. J.G.'s research was supported in part by the Perimeter Institute for Theoretical Physics. Research at Perimeter Institute is supported by the Government of Canada through Industry Canada and by the Province of Ontario through the Ministry of Research and Innovation.
Z.K. is supported in part by an Israel Science Foundation center for
excellence grant and by the I-CORE program of the Planning and Budgeting Committee
and the Israel Science Foundation (grant number 1937/12). Z.K. is also supported by the
ERC STG grant 335182 and by the United States-Israel BSF grant 2010/629.
H.O. is supported in part by
U.S.\ Department of Energy grant DE-SC0011632,
 by the Simons Investigator Award,
by the World Premier International Research Center Initiative,
MEXT, Japan,
by JSPS Grant-in-Aid for Scientific Research C-26400240,
and by JSPS Grant-in-Aid for Scientific Research on Innovative Areas
15H05895.
N.S. was supported in part by DOE grant DE-SC0009988. 
Y.W. was supported by the NSF grant PHY-1620059
and by the Simons Foundation Grant \#488653.
H.O. thanks the hospitality of the
Institute for Advanced Study and Harvard University,
where he spent his sabbatical in 2015  - 2016, and
of the Aspen Center for Physics, which is supported by
the National Science Foundation grant PHY-1066293. N.S. thanks the hospitality of  the Weizmann Institute of Science during the completion of this work.


  \appendix{A}{The Wess-Zumino Perspective}

Anomalies arise when we have some operator that should satisfy an operator equation, e.g. $\del^\mu j_\mu=0$ or $T_\mu^\mu=0$, but then one finds that such an equation is only correct at separated points
while at coincident points there are various contact terms. The essence is to show that these contact terms cannot be removed by redefining the scheme. Indeed, scheme redefinitions change the theory by various contact terms and so we need to demonstrate that the anomaly is invariant under scheme redefinitions. A convenient way to establish it is to introduce background fields for the various operators. Then scheme redefinitions correspond to adding new {\it local} terms to the action, which depend on these background fields and also, possibly, on the operators in the theory.

We would like to examine the operator $[\bar Q_+, \CO|]$, which is normally zero if $\CO$ is chiral. To this end we couple a background field to this operator.
A standard procedure is to couple the superfield $\CO$ to a background field in the superpotential but then we do not have a source for the redundant operator.
Therefore we will couple $\CO$ to a source in the K\"ahler potential. We add a corresponding term for a twisted chiral superfield $\CO$:
\eqn\gensources{\int d^4\theta \left(A \CO+B \tilde \CO+c.c.\right)~.}

Now we imagine computing the partition function $Z[A, B]$ (with $A,B$  superfields). It is useful to tabulate their charges
\eqn\chargei{\matrix{
& &U(1)_+ \times &U(1)_- & \hbox{Dimension} \cr
&A&       -1  &          -1 & (-1/2,-1/2)  \cr
&B&          -1  &          1 & (-1/2,-1/2)
}}

The standard expectation is that the partition function would not actually depend on most of the components in $A,B$ due the fact that $[\bar Q_{\pm}, \CO|]=0$ and $[\bar Q_+,\tilde \CO|]=[Q_-,\tilde \CO|]=0$.\foot{This standard expectation follows from the fact that these conditions hold ``off shell,'' namely there is a regularization where this is true. Technically, it means that there are no cohomologically nontrivial contact terms in correlation functions of these redundant operators.} So the standard expectation is that
\eqn\invari{Z[A, B]=Z[A+\bar D_+ \chi_-+\bar D_- \chi_+, B]~,}
\eqn\invarii{Z[A, B]=Z[A, B+\bar D_+ \psi_-+D_- \psi_+]~}
for arbitrary $\chi_{\pm},\psi_\pm$. This should be viewed, for example, in analogy with $\delta_\sigma Z[g_{\mu\nu}]=0$ for the conformal anomaly case.
What we would like to test is whether we can respect~\invari\ and \invarii\ while preserving ${\cal N}=(2,2)$ supersymmetry.
In other words, we want to see whether the shortening of the  background multiplets is consistent with supersymmetry.

The general principles that we reviewed above tell us that for infinitesimal $\chi_{\pm},\psi_\pm$,
$\delta_{\chi_\pm,\psi_\pm}Z[A, B]$
should be a  local functional of the sources and operators in the theory i.e.
\eqn\locvar{\delta_{\chi_\pm,\psi_\pm} \log Z[A, B]=\int \chi_- L^-_{local}+\cdots~,}
with $L_{local}$ some local function of the couplings and operators. The right hand side in~\locvar\ is restricted by demanding that it is supersymmetric and also  by demanding that it obeys the Wess-Zumino consistency conditions~\WessYU.

Let us assume that the partition function is invariant under the $\psi_{\pm}$ transformations, namely, $\tilde \CO$ obeys the twisted chiral shortening conditions at both separated and coincident points. We can then write the variation under $\chi_-$ as follows (the formula for the variation under $\chi_+$ is analogous)
\eqn\anoprop{\delta_{\chi_-} \log Z[A, B]=\kappa \int d^4\theta \bar D_+ D_- B \bar D_- \chi_-  {\cal J}_{++}~,}
with $\kappa$ some constant.
Equation~\anoprop\ respects supersymmetry (because it is a $\int d^4\theta$ integral), and it is consistent with the R-symmetry ($\bar D_+ D_- B$ carries R-charges $(0,0)$ and  $\chi_-$ carries R-charges $(-2,-1)$ and therefore $\bar D_- \chi_-$ carries $(-2,0)$ and thus it exactly cancels the $R$-charge of ${\cal J}_{++}$).
Furthermore,~\anoprop\ obeys the Wess-Zumino consistency condition since  $\bar D_+ D_- B$ is invariant under $B\to B+\bar D_+ \psi_-+D_- \psi_+$. Therefore,~\anoprop\ does not violate the fact that the partition function is postulated to be invariant under $\psi_\pm$ transformations.

We now proceed to prove that~\anoprop\ is cohomologically non-trivial. If~\anoprop\ were cohomologically trivial then one could add a local term to
$\log Z[A,B]$ such that the right hand side of~\anoprop\ would vanish while retaining supersymmetry and invariance under $\psi_{\pm}$ transformations.
It is clear (by integration by parts and using~\shortJa) that the $\chi_-$ variation of the local term \eqn\local{\int d^4\theta A\bar D_- D_-B  {\cal J}_{++}}
could cancel the right hand side of~\anoprop. However,~\local\ spoils the invariance of the partition function under $\psi_\pm$ transformations.
One can easily verify that indeed the right hand side of~\anoprop\ is physical as long as we insist on supersymmetry and invariance under $\psi_\pm$ transformations.

To summarize let us make some comments 

\medskip

\item{1.} Suppose we always insist on preserving $\cN=(2,2)$ supersymmetry. Then, if $\kappa$ in~\anoprop\ is nonzero, then it turns out that we may not be able to respect both~\invari\ and \invarii. In other words, we cannot embed the coupling constants of chiral and twisted chiral operators into short multiplets. At least some of the couplings have to be in longer multiplets.
\item{2.} We can view the $\chi$ and $\psi$ transformations as analogous to $U(1)_A$ and $U(1)_V$ transformations in $2d$ electrodynamics. If we preserve one we must give up on the other, but choosing which one to preserve is at our discretion. Therefore, the situation is very similar to the way the usual chiral 't Hooft anomalies arise~\FrishmanDQ.

\medskip

Note that from equation~\anoprop\ we can immediately write the anomaly in operatorial formalism. This is because $\chi_-$ couples to $\bar D_+\CO$ and so we find
\eqn\operatorial{\bar D_+ \CO \sim \kappa (\bar D_- \bar D_+ D_- B) {\cal J}_{++}~.}
But since our partition function is invariant under $\psi_\pm$ transformations and hence depends only on $\tilde\lambda=\bar D_+ D_- B$, which is a standard twisted chiral background field, we can also write
\eqn\operatoriali{\bar D_+ \CO \sim \kappa \bar D_- \tilde \lambda\ {\cal J}_{++}~.}
Hence in a ``fermionic background'' for the twisted chiral coupling, the operator $\CO$ ceases to be chiral.

It is now straightforward to make contact with the analysis in the bulk of the paper. Our discussion in this appendix has shown that there may be an anomaly with coefficient $\kappa$ and that it would manifest itself as~\operatoriali. Comparing with~\operatorialiia\ we thus see that this coefficient is nonzero whenever the OPE coefficient in~\contactOPE\ is nonzero.
Furthermore, the analysis in this appendix sheds light on the choices we could make in~\cancelAA. Indeed, we could have chosen whether to postulate that the partition function preserves~\invari,\invarii, or none of the two. As in all cases with anomalies, these various choices are related to each other by adding counter-terms to the action, e.g. the one we discussed in~\local. By choosing the coefficient of this counter-term carefully, we can change the scheme from the one where $\psi$ transformations are obeyed to the one where $\chi$ transformations are obeyed.

\appendix{B}{Special Geometry Relation}
It is well known that for an $\cN=(2,2)$ SCFT, the Riemannian curvature for marginal deformations generated by chiral primary fields
is determined in terms of the Zamolodchikov metric and the chiral ring coefficients  \refs{\Strominger,\ttstar,\BershadskyCX,\BaggioIOA}.
For $c=9$, this is known as the special geometry relation. Below we will give a simple derivation of this relation using \form\ (our derivation is valid for any central charge $c$).

Following \DixonFJ, we can use $\cN=2$ Ward identities to express the integrand of \form\ as a total derivative,
\eqn\curvtwo{\eqalign{
	&\vev{ F_i(1)  \bar F_{\bar j}(y) F_{ k} (\infty) \bar F_{\bar \ell}(0)}_c\cr
	=
	&
	\pa_{y^{++}} \pa_{y^{--}}\pa_{w^{++}} \pa_{w^{--}}  \left(
	 (y^{++}-w^{++})( y^{--}-w^{--})
	\la \CO_{i}(1)  \bar\CO_{\bar j}(y)  \CO_{k}(\infty) \bar\CO_{\bar \ell}(w)\ra_c
	\right) \biggr|_{w=0}~.}}
The curvature is then computed by integration by parts, with non-vanishing contributions coming from  the origin ($y=0$).\foot{A priori, there can be boundary contributions from contours around $y=1$ and $\infty$ as well. However careful analysis of the integrand shows that such contributions are absent for this particular four-point function.
} Therefore,
we need to study the order-one terms in the limit $\lim_{y\rightarrow 0}\pa_{w^{++}} \pa_{w^{--}}  \left(
(y^{++}-w^{++})( y^{--}-w^{--})
\la \CO_{i}(1)  \bar\CO_{\bar j}(y)  \CO_{k}(\infty) \bar\CO_{\bar \ell}(w)\ra_c
\right) \biggr|_{w=0}$. We can act with the derivatives on the prefactor $|y-w|^2$ and thus reduce the problem to studying the
$y\rightarrow 0$ limit of $\la \CO_{i}(1)  \bar\CO_{\bar j}(y)  \CO_{k}(\infty) \bar\CO_{\bar \ell}(0)\ra_c$.
There is a contribution from the unit operator in the $t$ and $u$ channel as well as a contribution from the   $(2,2)$ chiral
primaries,
\eqn\SG{R_{k\bar \ell i \bar j }   =- \lim_{y\rightarrow 0}  \la \CO_{i}(1)  \bar\CO_{\bar j}(y)  \CO_{k}(\infty) \bar\CO_{\bar \ell}(0)\ra_c
	=g_{i\bar j} g_{k\bar \ell}+g_{i\bar \ell} g_{k\bar j }  -\bar C_{\bar  j \bar \ell }{}^{\bar J}g_{I\bar J}C_{ik}{}^{I} ~, }
where  $g_{I\bar J}$ is the metric associated with the R-charge $(2,2)$ chiral primaries.

\appendix{C}{Four-Point Function involving Supercurrents}

Let us focus on the connected four-point function
\eqn\scfp{
\la
S^{\A A}_{+++}(x)\,F^{C\dot C}_i(y)\,F^{D\dot D}_j(z)  S^{ \B B}_{+++}(w)
\ra_c \,.
}
Since the four-point function is holomorphic in $z$ and $w$, it is determined completely by the poles in the OPE between the supercurrents and other insertions. We first consider the singularities in $x$, and denote the polar terms in $(x-w), (x-y),( x-z)$ by $I_1, I_2, I_3$ respectively.

From the OPE of the $\cN=4$ supercurrents,
\eqn\SSOPE{
S_{+++}^{\A A}(x) S_{+++}^{\B B}(w)=
\epsilon^{AB}\left({4k \epsilon^{\A\B}\over  (x-w)^3}
-{4 \sigma_i^{\A\B} J^i(w) \over (x-w)^2}
+{2 \epsilon^{\A\B} T( w)-2 \sigma_i^{\A\B} \pa J^i(w) \over (x-w) }
\right)+\dots \,,
}
we have
\eqn\Ia{
I_1=\left\langle
{2 \epsilon^{AB} \epsilon^{\A\B} T( w) \over (x-w) }
F_i^{C\dot C} (y) F_j^{D\dot D}(z)
\right\rangle \,,
}
where we have dropped the disconnected pieces and also the terms involving the $SU(2)_R$ currents since $F_i$ are $SU(2)_R$ singlets. It is then easy to obtain using the OPE between $T$ and $F_i$
\eqn\Iaa{
I_1={2 \epsilon^{\A\B}\epsilon^{AB}\epsilon^{CD}\epsilon^{\dot C\dot D}\eta_{ij}\over (x-w)(w-y)^2(w-z)^2(\bar y-\bar z)^2}\,.
}
Similarly looking at the OPE between $S_{+++}^{\A A}$ and $F_i^{C\dot C}$
\eqn\SFO{
	S_{+++}^{\A A }(z) F^{B\dot B}(w)= {1\over 2}\pa_{w^{++}}\left(
	{\epsilon^{\A \B}\epsilon^{AB}  Q^{\dot \B\dot B}_-\CO_{\B\dot \B}\over z^{++}-w^{++}},
	\right)+\dots
}
we have
\eqn\Ib{
I_2=
{1\over 2}
\la
\pa_{y^{++}}\left(
{\epsilon^{\A \C}\epsilon^{AC}  Q^{\dot \C \dot C}_-\CO_{\C\dot \C}(y)\over  x-y}
\right)
F_j^{D\dot D}(z) S^{\B B}_{+++}(w)
\ra \,.
}
Using again \SFO~and also
\eqn\SO{
	S_{+++}^{\A A }(z) \CO_{ \B \dot \B}(w)={\D^\A_\B  \over 2(z^{++}-w^{++})} Q_+^{\C A} \CO_{ \C \dot \B}(w)+\dots \,,
}
we get
\eqn\Ibb{
I_2={2\epsilon^{\A\B} \epsilon^{AC} \epsilon^{BD}\epsilon^{\dot C\dot D} \eta_{ij}(2y-x-w) \over (\bar z-\bar y)^2(y-w)^2(z-w)^2(x-y)^2 }.
}
Finally from the OPE between $S_{+++}^{\A A}$ and $F_j^{D\dot D}$ we get
\eqn\Ic{
I_3=
{1\over 2}
\left\la
F_i^{C\dot C}(y)  \pa_{z^{++}}\left(
{\epsilon^{\A \D}\epsilon^{AD}  Q^{\dot \D \dot D}_-\CO_{\D\dot \D}(z)\over  x-z}
\right)
S^{\B B}_{+++}(w)
\right\ra~,
}
and following the same steps as above we obtain
\eqn\Icc{
I_3={ 2\epsilon^{\A\B} \epsilon^{AD} \epsilon^{BC}\epsilon^{\dot C\dot D} \eta_{ij}(w+x-2z) \over (\bar z-\bar y)^2(y-w)^2(z-w)^2(x-z)^2 }.
}
Putting together \Iaa, \Ibb~and \Icc~while taking the limit $w\to \infty$, we arrive at
\eqn\ssfpfinal{\eqalign{
\la
&S^{\A A}_{+++}(0)\,F^{C\dot C}_i(y)\,F^{D\dot D}_j(1)  S^{ \B B}_{+++}(\infty)
\ra_c
\cr
=&  2\epsilon^{\A\B}\epsilon^{\dot C\dot D} \eta_{ij}
\left(
{ -\epsilon^{AC} \epsilon^{BD}\over  y^2(\bar y-1)^2 }+{ \epsilon^{AD} \epsilon^{BC}\over (\bar y-1)^2}
\right)
\cr
=&
-2\epsilon^{\A\B}\epsilon^{\dot C\dot D} \eta_{ij}
\pa_{y^{++}}  \pa_{y^{--}} \left(
{\epsilon^{AC} \epsilon^{BD}\over  y (\bar y-1) }+{ \epsilon^{AD} \epsilon^{BC} y\over \bar y-1}
\right).
}}

\appendix{D}{ $SU(2)_{{\rm out}}$ Selection Rules}
In this appendix we prove that
\eqn\outward{\eqalign{
		&  \vev{F_{i1\dot 1}(0)\,F_{k1\dot 1} (y)\,  F_{\ell 2\dot 2} (1)\,F_{j2\dot 2} (\infty )}+\vev{F_{i2\dot 1}(0)\,F_{k1\dot 1} (y)\,  F_{\ell 1\dot 2} (1)\,F_{j2\dot 2} (\infty )}
		\cr
		&
		+\vev{F_{i2\dot 1}(0)\,F_{k1\dot 1} (y)\,  F_{\ell 2\dot 2} (1)\,F_{j1\dot 2} (\infty )}=0  ,		
	} }
follows	from $\cN=(4,4)$ superconformal Ward identities. More generally, the four-point function of the exactly marginal operators $  \vev{F_{i A \dot A}F_{j B\dot B}F_{k C\dot C}F_{\ell D\dot D}}$ respects $SU(2)_{{\rm out}}$ Ward identities.\foot{The $SU(2)_{{\rm out}}$ is the outer automorphism group of the ${\cal N}=4$ superconformal algebra, but not a symmetry of the full SCFT. However, the operator spectrum of the theory can still be organized into $SU(2)_{{\rm out}}$ representations  by postulating that the $\cN=4$ superconformal primaries transform as singlets. Here we see that certain correlation functions respect the $SU(2)_{\rm out}$ invariance. This is analogous to the bonus $U(1)_Y$ symmetry in the $4d$ $\cN=4$ super Yang-Mills theory \IntriligatorFF.  More recently, the outer automorphism group figured in the study of the K3 SCFT from spacetime arguments in \LinDSA. 	
	 }
	
	For notational simplicity, let us focus on the left-moving side of \outward. Using $\cN=4$ Ward identities
	\eqn\FFa{\eqalign
		{
			&\vev{
				Q^{2  2}_{+} \CO_i^1(x) Q^{2  2}_{+} \CO_k^1(y) Q^{1  1}_{+} \CO_\ell^2(z) Q^{1  1}_{+} \CO_j^2(\infty ) }
			\cr=
			&2\pa_{y^{++}}\vev{
				Q^{2  2}_{+} \CO_i^1(x)   \CO_k^1(y)  \CO_\ell^2(z) Q^{1  1}_{+} \CO_j^2(\infty ) }
			-
			2\pa_{x^{++}}\vev{
				\CO_i^1(x)   	Q^{2  2}_{+}\CO_k^1(y)  \CO_\ell^2(z) Q^{1  1}_{+} \CO_j^2(\infty ) },
		}}
		\eqn\FFb{\eqalign
			{
				&\vev{
					Q^{2  1}_{+} \CO_i^1(x) Q^{2  2}_{+} \CO_k^1(y) Q^{1  2}_{+} \CO_\ell^2(z) Q^{1  1}_{+} \CO_j^2(\infty ) }
				\cr=
				&	
				2\pa_{x^{++}}\vev{
					\CO_i^1(x)   	Q^{2  2}_{+}\CO_k^1(y)  \CO_\ell^2(z) Q^{1  1}_{+} \CO_j^2(\infty ) }
				-\vev{
					Q^{2  1}_{+}	\CO_i^1(x)  	Q^{2  2}_{+} \CO_k^1(y)  \CO_\ell^2(z)  \CO_j^1(\infty ) },
						}}
			and
			\eqn\FFc{\eqalign
				{
					&\vev{
						Q^{2  1}_{+} \CO_i^1(x) Q^{2  2}_{+} \CO_k^1(y) Q^{1  1}_{+} \CO_\ell^2(z) Q^{1  2}_{+} \CO_j^2(\infty ) }
					\cr=
					&	
					2\pa_{y^{++}}\vev{
						Q^{2  1}_{+}	\CO_i^1(x)   \CO_k^1(y)  \CO_\ell^2(z) Q^{1  2}_{+} \CO_j^2(\infty ) }
					+\vev{
						Q^{2  1}_{+}	\CO_i^1(x)  	Q^{2  2}_{+} \CO_k^1(y)  \CO_\ell^2(z)  \CO_j^1(\infty ) } .
				}}
				Moreover, from \SFO\ we can derive
				\eqn\FFd{\eqalign	{
						\vev{
							Q^{2  1}_{+}	\CO_i^1(x)   \CO_k^1(y)  \CO_\ell^2(z) Q^{1  2}_{+} \CO_j^2(\infty ) }
						=-\vev{
							Q^{2  2}_{+}	\CO_i^1(x)   \CO_k^1(y)  \CO_\ell^2(z) Q^{1  1}_{+} \CO_j^2(\infty ) } .
					}}		
					Putting together \FFa, \FFb, \FFc~and \FFd, we obtain
					\eqn\FFI{\eqalign
						{
							&\vev{
								Q^{2  2}_{+} \CO_i^1(x) Q^{2  2}_{+} \CO_k^1(y) Q^{1  1}_{+} \CO_\ell^2(z) Q^{1  1}_{+} \CO_j^2(\infty ) }
							+\vev{
								Q^{2  1}_{+} \CO_i^1(x) Q^{2  2}_{+} \CO_k^1(y) Q^{1  2}_{+} \CO_\ell^2(z) Q^{1  1}_{+} \CO_j^2(\infty ) }
							\cr
							&	+
							\vev{	Q^{2  1}_{+} \CO_i^1(x) Q^{2  2}_{+} \CO_k^1(y) Q^{1  1}_{+} \CO_\ell^2(z) Q^{1  2}_{+} \CO_j^2(\infty ) }
							=0
					,	}}
						which leads to \outward.

 \listrefs

\bye